\documentclass[conference]{IEEEtran}
\usepackage{url}
\usepackage{enumitem,kantlipsum}
\usepackage{subfig}
\usepackage{array}
\usepackage{amsmath,amssymb,amsfonts}
\usepackage{graphicx}
\usepackage{textcomp}
\usepackage{xcolor}
\usepackage{caption}
\usepackage{multirow}
\usepackage{pifont}
\usepackage{colortbl}
\usepackage{wasysym}
\usepackage{fontawesome}
\usepackage{comment}

\usepackage{booktabs, siunitx}
\usepackage{color, colortbl}
\definecolor{Gray}{gray}{0.9}
\definecolor{LightCyan}{rgb}{0.88,1,1}
\definecolor{green}{HTML}{3049D4}
\definecolor{codepurple}{rgb}{0.58,0,0.82}
\newcolumntype{C}[1]{>{\centering\arraybackslash}p{#1}}

\usepackage[linesnumbered,ruled,noline]{algorithm2e}

\SetCommentSty{mycommfont}
\SetKwInput{KwInput}{Input}                
\SetKwInput{KwOutput}{Output}              
\SetKwBlock{Loop}{Loop}{}
\SetKwBlock{If}{if}{}
\SetKwBlock{Func}{Function}{}

\newcommand{\ours}{\textsf{Morph}\xspace}
\newcommand{\red}[1]{{\color{black} #1}}
\newcommand{\blue}[1]{{\color{black} #1}}

\begin{document}


\title{Morph: ChirpTransformer-based Encoder-decoder Co-design for Reliable LoRa Communication}

\author{
\IEEEauthorblockN{Yidong Ren$^{1}$, Maolin Gan$^{1}$, Chenning Li$^{2}$, Shakhrul Iman Siam$^{3}$, Mi Zhang$^{3}$, Shigang Chen$^{4}$, Zhichao Cao$^{1}$}
\IEEEauthorblockA{$^{1}$Michigan State University, $^{2}$Massachusetts Institute of Technology, $^{3}$Ohio State University, $^{4}$University of Florida\\
\{renyidon, ganmaoli, caozc\}@msu.edu, lichenni@mit.edu, \{siam.5, mizhang.1\}@osu.edu, sgchen@cise.ufl.edu}}





\maketitle

\begin{abstract}

In this paper, we propose \ours, a LoRa encoder-decoder co-design to enhance communication reliability while improving its computation efficiency in extremely-low signal-to-noise ratio (SNR) situations.
The standard LoRa encoder controls 6 Spreading Factors (SFs) to tradeoff SNR tolerance with data rate. SF-12 is the maximum SF providing the lowest SNR tolerance on commercial off-the-shelf (COTS) LoRa nodes. In \ours, we develop an SF-configuration based encoder to mimic the larger SFs beyond SF-12 while it is compatible with COTS LoRa nodes.
Specifically, we manipulate four SF configurations of a \ours symbol to encode 2-bit data.
Accordingly, we recognize the used SF configuration of the symbol for data decoding. We leverage a Deep Neural Network (DNN) decoder to fully capture multi-dimensional features among diverse SF configurations to maximize the SNR gain. Moreover, we customize the input size, neural network structure, and training method of the DNN decoder to improve its efficiency, reliability, and generalizability. 
We implement \ours with COTS LoRa nodes and a USRP N210, then evaluate its performance on indoor and campus-scale testbeds. 
Results show that we can reliably decode data at -28.8~dB SNR, which is 6.4~dB lower than the standard LoRa with SF-12 chirps. In addition, the computation efficiency of our DNN decoder is about 3$\times$ higher than state-of-the-art.

\end{abstract}

\section{Introduction}
\label{sec-introduction}

LoRa has emerged as a promising mechanism to connect unattended Internet-of-Things (IoT) devices at scale.
Standard LoRa physical layer (LoRa-PHY) adopts Chirp Spread Spectrum (CSS) modulation to meet the envisioned long-distance communication, allowing successful decoding even below the noise floor~\cite{sun2023flora,li2022lora, ren2023prism}.
Spreading Factor (SF) configurations determine the duration of a chirp. The larger a chirp's SF is, the lower the signal-to-noise ratio (SNR) requirement is for decoding.
The noise tolerance is enhanced when SF increases at a lower data rate or energy efficiency cost. Thus, a LoRa node's SF will be configured to the smallest one ensuring that data can be reliably decoded under its SNR condition while reserving energy efficiency.

SFs can be scaled from 7 to 12 on commercial off-the-shelf (COTS) LoRa nodes, resulting in a communication range of 9.08-30~km in line-of-sight (LoS) conditions~\cite{liando2019known, petajajarvi2015coverage, ren2025toward}.
However, in a wide range of IoT applications like forest/urban environment surveillance~\cite{greenorbs, citysee, ren2025toward, ren2025aeroecho}, tunnel monitoring~\cite{tunnel}, industrial IoT~\cite{sanchez2016state}, and urban shared bicycle tracking~\cite{Losee, deepLora}, LoRa nodes are usually blocked by surrounding obstacles such as buildings and trees (i.e., non-line-of-sight (NLoS) scenarios)~\cite{petajajarvi2015coverage, liando2019known, deepLora, lin2024multi}. 
Consequently, the observed 100~m-2.6~km NLoS communication range~\cite{liando2019known, bor2016lora, wixted2016evaluation, centenaro2016long, iova2017lora, kartakis2016demystifying, deepLora} is much shorter than the expected (5~km or above)~\cite{LoRaWAN}. 
%
%
The root cause is that compared to LoS scenarios, the blockage in those NLoS scenarios easily makes the SNR lower than the threshold that the largest SF-12 can tolerate.
A recent measurement study~\cite{losee22ren} has shown that we can increase the coverage area from 11.4~km$^2$ to 15.2~km$^2$ with 2~dB SNR gain in an urban environment.
Hence, to realize reliable LoRa communication in practice, the gap between the possible extremely-low SNR in NLoS scenarios and the minimum SNR that LoRa-PHY can tolerate needs to be bridged.

%
Some methods~\cite{dongare_charm_2018,eletreby_empowering_2017,gadre_frequency_nodate,li2021nelora,hou2022malora} redesign the physical layer decoder to obtain SNR gain. 
NELoRa~\cite{li2021nelora,du2023nelora} has developed a neural-enhanced decoder utilizing a Deep Neural Network (DNN) model that captures multi-dimensional features from both amplitude and phase spectrograms to lower the requested SNR of reliable decoding compared with the standard LoRa decoder. Charm~\cite{dongare_charm_2018} and MALoRa~\cite{hou2022malora} utilize multiple gateways or antennas to enhance the received signal strength by coherently combining the duplicate signals during decoding.
These methods obtain SNR gain with a tradeoff of computation efficiency and infrastructure cost.
However, when SNR is getting lower and a larger SF is adopted, the SNR gain of these methods has to be compromised by the exhausting DNN computation overhead~\cite{li2021nelora} and the residual phase-offset among multiple signal copies~\cite{hou2022malora}. 
%

\begin{figure}[t!]
\centering
\includegraphics[width=0.95\linewidth]{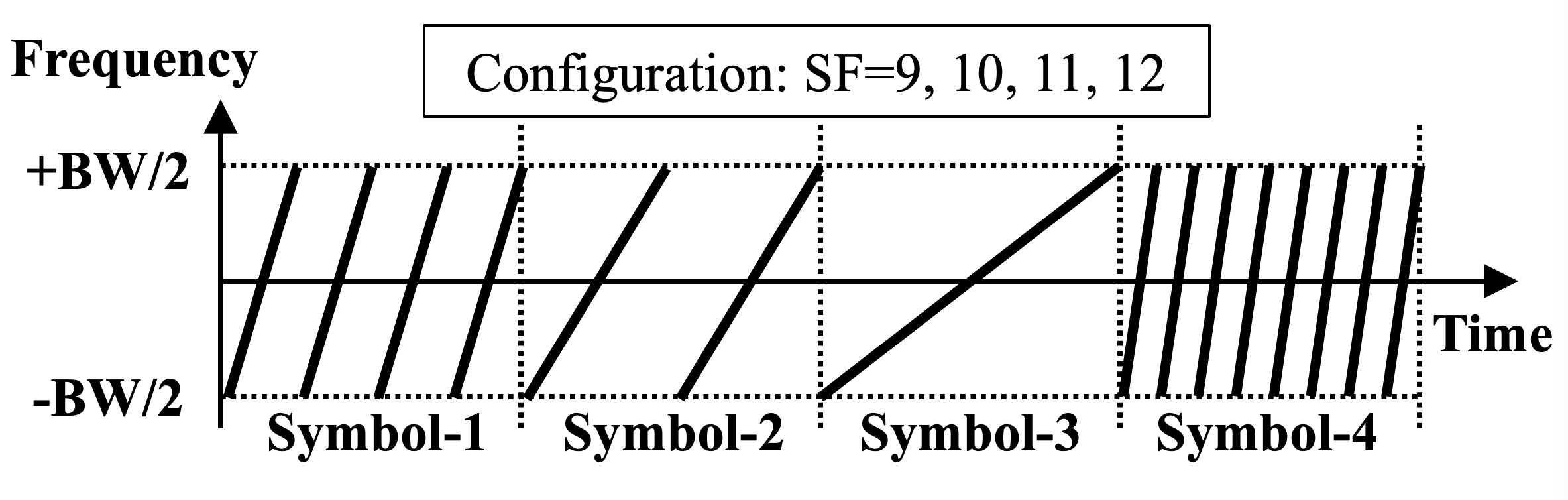}
\vspace{-3mm}
\caption{The illustration of SF-configuration-based symbol encoding, using four SFs (e.g., 9-12) to represent 2-bit data.}
\vspace{-3mm}
\label{fig:cover}
\end{figure}

%
In this paper, we propose \ours, a LoRa encoder-decoder co-design that achieves reliable and efficient symbol decoding at the extremely-low SNR with commercial-of-the-shelf (COTS) LoRa nodes. 
We rebuild a compatible encoder based on ChirpTransformer~\cite{ren2024chirptransformer} to tolerate extremely-low SNR inherently, just like the unavailable larger SFs beyond SF-12 do. The symbol signals generated by the encoder can be efficiently decoded by a neural-enhanced decoder for SNR gain maximization. 
%
%
As illustrated in Figure~\ref{fig:cover}, \ours incorporates an SF-configuration-based encoder that uses the base up-chirps across four SF configurations (e.g., 9-12) to form four symbols to encode 2-bit data. These four symbols have the same period that equals the duration of the maximum SF chirp (e.g., 12). We decode a symbol by recognizing the SF configuration of the symbol.
Different symbols exhibit diverse time-domain patterns, which can benefit \ours decoding at extremely-low SNR. 
In practice, we use the hardware interrupt that triggers frequency hopping~\cite{SX127X} to enable the envisioned SF switching between two adjacent symbols of a \ours packet on COTS LoRa nodes.

At the decoder side, we design a Deep Neural Network (DNN)-based decoding method to enhance the SNR tolerance further while optimizing the computation efficiency. For the 4-class pattern recognition problem, the neural-enhanced decoder can obtain extra SNR gain by fully capturing the time-domain features among different SF configurations. We customize the DNN model in three folds. 
First, considering the overwhelming noise at extremely-low SNR, we compress the input spectrograms to achieve the best balance between symbol feature reservation and noise reduction. 
Second, targeting the 4-class classification problem, we optimize the DNN structure to improve the computational efficiency while keeping the decoding accuracy for practical deployments.
Third, we develop a few-shot learning strategy with initial phase augmentation to achieve a ``one-fits-all'' DNN model with a low training overhead. Thus, the DNN decoder can adapt well to diverse environments and hardware.

We have implemented \ours on COTS LoRa nodes and a USRP N210 as the signal recorder connected to a local server or a Raspberry Pi for decoding. Given our indoor and campus-scale ($2800m\times1700m$) testbeds, we conduct extensive experiments to evaluate \ours.
The results show that \ours enables successful data decoding at -28.8~dB SNR, which is 6.4~dB lower than LoRa-PHY with SF-12 chirps, thus significantly reducing the symbol error rate (SER) of those originally unreachable LoRa links with LoRa-PHY to 40.54\% on average in our outdoor deployment. 
Moreover, the computation efficiency of our DNN decoder is $3.14\times$ higher than state-of-the-art NELoRa~\cite{li2021nelora}.
%

%
In summary, our contributions are listed as follows: 
\begin{itemize}[leftmargin=*]
    \item 
    We propose a novel LoRa encoder design to enable reliable LoRa communication inherently while keeping computation friendly to a neural-enhanced decoder. The encoder is compatible with COTS LoRa nodes.
    \item 
    For enhancing \ours's efficiency and generalization in real-world deployments, several adaptive DNN customization techniques are involved, which can be adopted by other neural-enhanced wireless systems.  
    \item 
    We implement a prototype of \ours using COTS LoRa nodes and a USRP. 
    The results show that \ours can tolerate SNR as low as -28.8~dB and significantly improve the computation efficiency of DNN decoding.
\end{itemize}

\section{Background and Motivation}
\label{sec-background-motivation}

\begin{figure}[t!]
\centering
\includegraphics[width=1\linewidth]{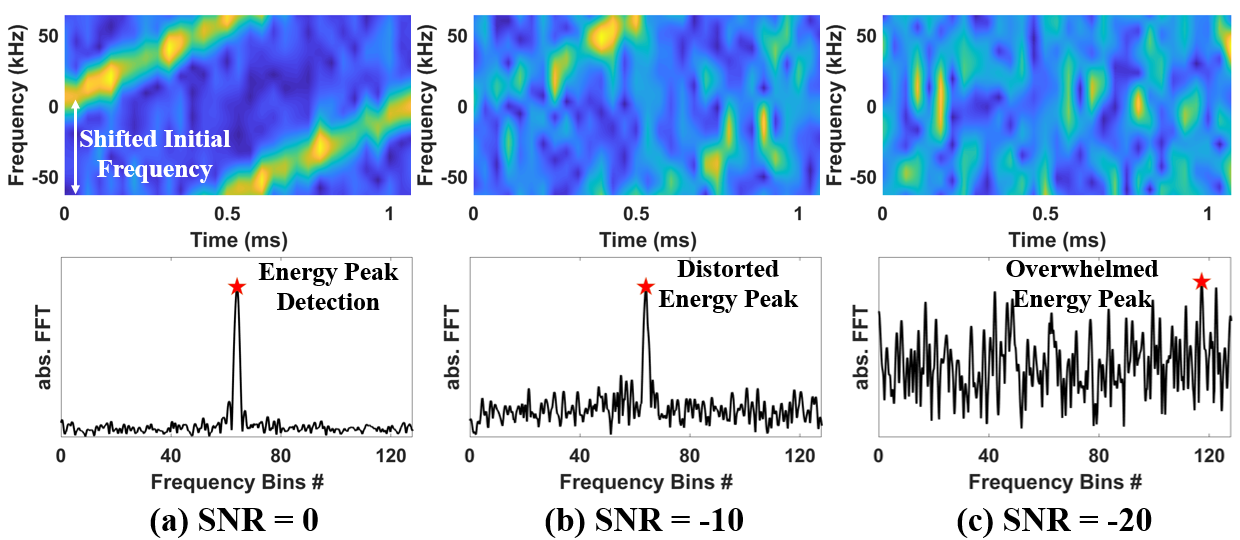}
\vspace{-4mm}
\caption{The illustration of LoRa-PHY encoding and decoding. (a) shifted initial frequency encoding (top) and energy peak detection based decoding (bottom). (b) and (c) the dechirp suffers as SNR decreases.}
\label{fig:dechirp-processing}
\vspace{-4mm}
\end{figure}

\subsection{Encoding and Decoding in LoRa-PHY}
\label{subsec-lora-phy}

We start by briefly describing the standard encoder/decoder used in LoRa-PHY~\cite{eletreby_empowering_2017, li2021nelora}. 
At the encoder side, LoRa-PHY uses bandwidth ($BW$) to configure a base up-chirp symbol, whose frequency increases linearly from $-\frac{BW}{2}$ to $\frac{BW}{2}$ over time. Given the base up-chirp, data bits are encoded by shifting the initial frequency of a base up-chirp to $f_{s}$, shown at the top of Figure~\ref{fig:dechirp-processing}.
At the decoder side, dechirp~\cite{eletreby_empowering_2017, li2021nelora} is the standard decoding process in LoRa-PHY.
First, the received chirp symbol is multiplied with a base down-chirp (i.e., the conjugate of a base up-chirp). The Fast Fourier Transform (FFT) is then used to focus on the energy of the chirp symbol at a single tone of $f_s$ on the spectrum~\cite{eletreby_empowering_2017,tong_combating_2020}. As such, data bits can be decoded by detecting the spectral energy peak as shown at the bottom of Figure~\ref{fig:dechirp-processing}(a). 
For successfully decoding a chirp symbol under low SNR conditions, the energy peak should be higher than the noise energy peak in the spectrum~\cite{li2021nelora}. 

To illustrate the negative impact of noise, we decrease the SNR of the same chirp symbol to -10~dB and -25~dB by adding Gaussian white noise with controlled amplitude and phase~\cite{tong_combating_2020, tong_colora_2020}. Figure~\ref{fig:dechirp-processing}(b) and (c) show the spectrograms of the interfered chirp symbols and the corresponding spectrum derived by dechirp. From the spectrum, we can see the energy peak is distorted and even overwhelmed by noise energy. Thus, to successfully decode a chirp symbol, the energy peak should be higher than the noise energy peak in the spectrum. 

\subsection{SNR Threshold and SF-12 Beyond}
\label{subsec-snr-threshold}

\blue{In LoRa-PHY, SF is a configurable parameter that determines the period of a chirp symbol and data rates.} Specifically, a chirp symbol consists of $2^{\text{SF}}$ chips, which encode SF bits where each chip lasts $\frac{1}{BW}$ second. Therefore, a chirp symbol with larger SF has a longer duration and can tolerate lower SNR because more energy is accumulated at frequency $f_s$ through dechirp.
To empirically measure the required SNR of the dechirp under different SFs, we generate $10\times 2^{\text{SF}}$ chirp symbols for each configuration across different SF and SNR under $BW$ 125~kHz. 
Specifically, we collect approximately 100,000 chirp symbols through real wireless channels using our indoor testbed ($\S$\ref{sec_implementation}) with 6 SF configurations (i.e., 7-12) using 125~kHz $BW$ at SNR of 32-37~dB. 
We add Gaussian white noise with controlled amplitudes to the collected I and Q traces~\cite{tong_colora_2020,tong_combating_2020}, rendering many chirp symbols at SNR of -40-20~dB.
We apply dechirp to decode the emulated chirp symbols and calculate the SER at different SNRs.

\begin{figure}[!t]
    \centering
\vspace{-4.5mm}
    \subfloat[Dechirp SER]{
    \label{fig:theoretical-limit-1}
    \includegraphics[width=0.24\textwidth]{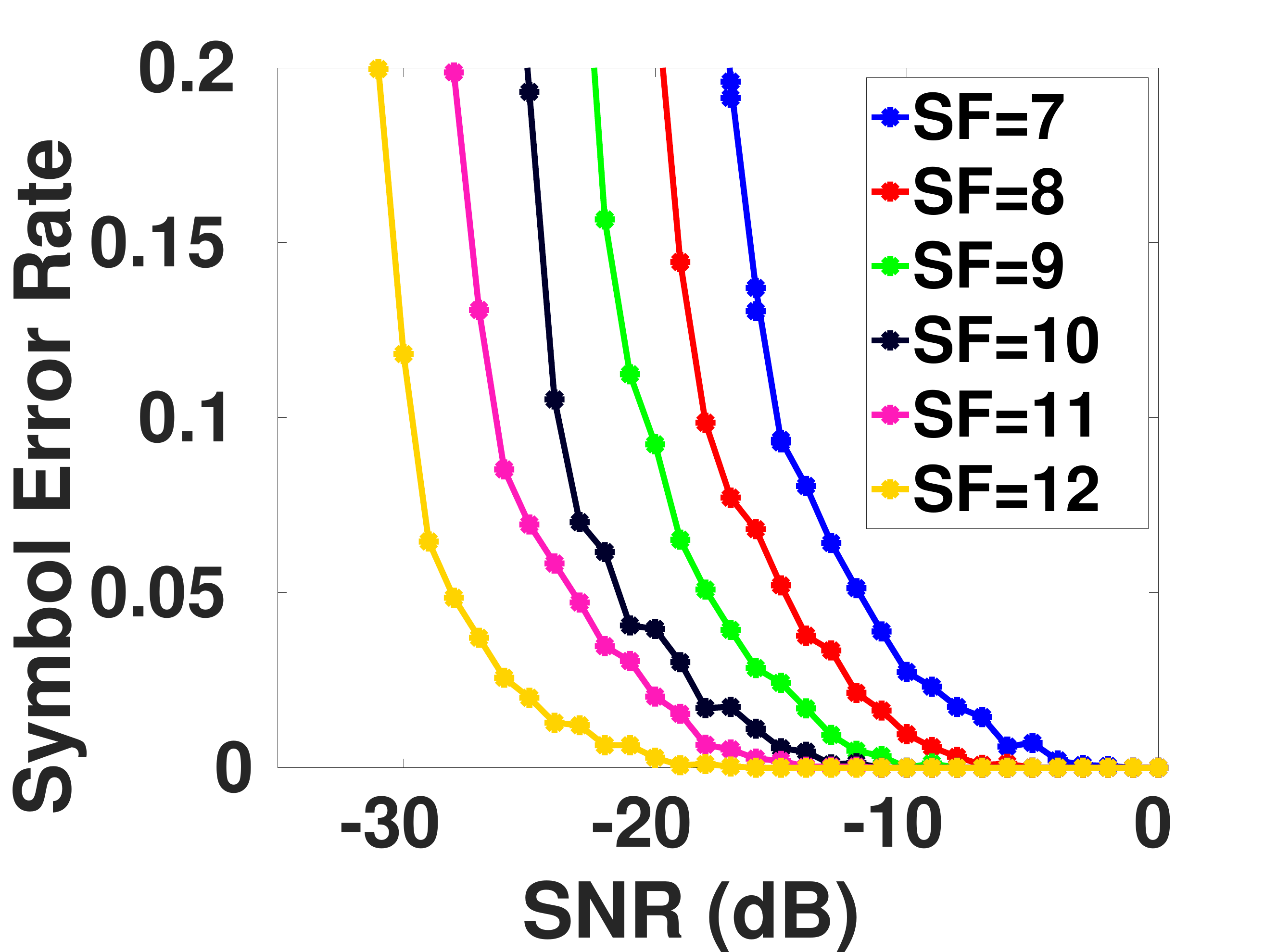}
    }
    \subfloat[SNR Threshold]{
    \label{fig:theoretical-limit-2}
    \includegraphics[width=0.22\textwidth]{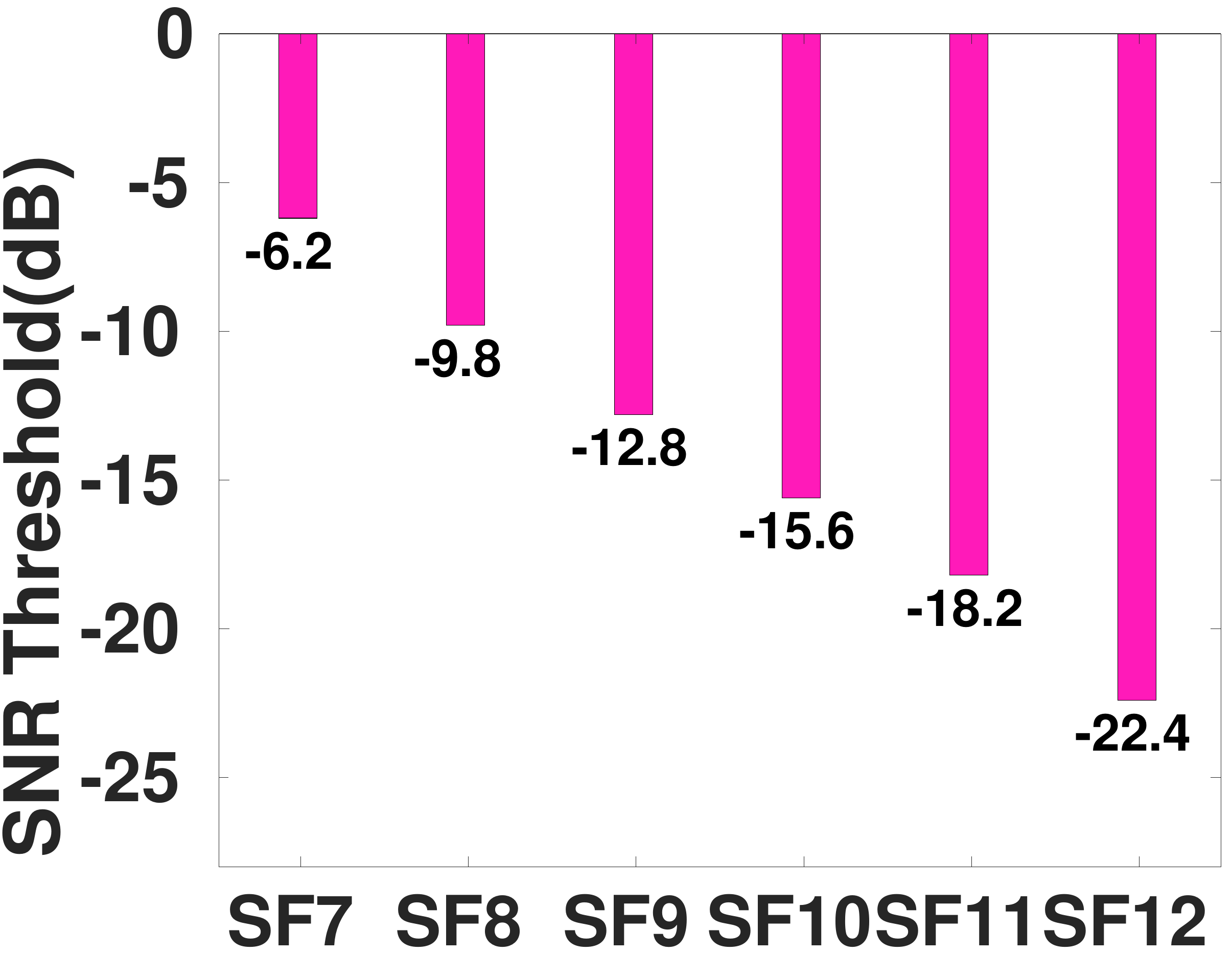}
    }
    \vspace{-1mm}
    \caption{The influence of SF configurations on SNR tolerance. (a) SER distribution under different SNR. (b) SNR threshold.}
    \label{fig:theoretical-limit}
\vspace{-3mm}
\end{figure}

Figure~\ref{fig:theoretical-limit-1} shows the SER distribution with different SF configurations. 
We can see that dechirp can achieve better SNR tolerance on chirp symbols with a larger SF.
Figure~\ref{fig:theoretical-limit-2} further shows the SNR threshold of different SF configurations, above which dechirp can decode with less than 1\% SER.
The results show that increasing SF by one can achieve a 2.6-4.2~dB gain of the SNR threshold. Dechirp achieves the minimum -22.4~dB SNR threshold with the SF-12 chirp symbols.
\blue{Ideally, we can tolerate lower SNR condition by increasing SF and reduce date rates at the encoder side. However, the maximum SF is 12 and the minimum forward error correction coding rate is 4/8 for COTS LoRa nodes.} With SF-12 configuration, a recent LoRa measurement study~\cite{losee22ren} has shown that although the longest communication range can reach 3.2~km - 3.5~km, a gateway can only cover about an irregular 11~km$^2$ - 12~km$^2$ area in an urban environment, which is far from needed. Meanwhile, the study has observed that a 2~dB SNR gain can increase the coverage area to about 15~km$^2$. \blue{This motivates us to develop a novel encoding scheme for COTS LoRa nodes. Our goal is to enable reliable communication at lower SNR levels than the limited threshold provided by existing LoRa-PHY in SF-12 chirps.}

\section{Key Design Overview}
\label{sec-key-design-choices}


In this section, we introduce the underlying principle and tradeoff behind the design of \ours before presenting details in the next section.

\begin{figure}[t!]
\centering
\includegraphics[width=1\linewidth]{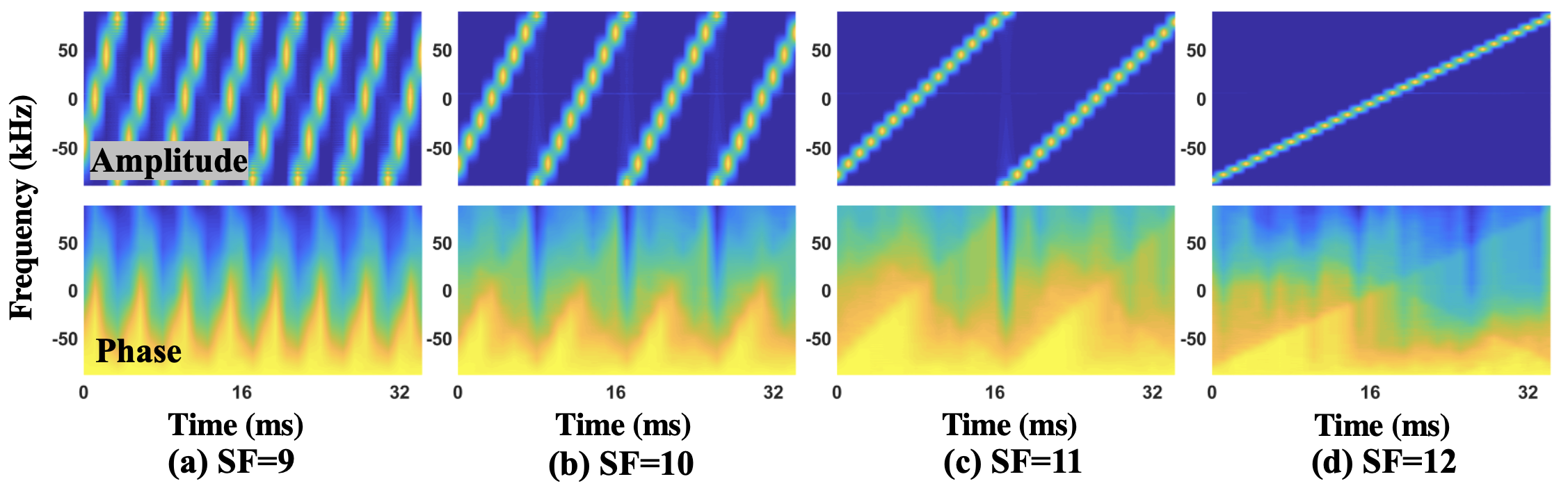}
\caption{The amplitude and phase spectrograms of continuous base up-chirps with different SF configurations in the period of an SF-12 chirp.}
\label{fig:spectrogram_illustration}
\vspace{-3mm}
\end{figure}

\noindent
\textbf{Low-SNR and COTS-compatible Encoder:} According to Shannon–Hartley theorem~\cite{hartley1928transmission, shannon1949communication}, given fixed channel bandwidth, it is a tradeoff between data rate and SNR tolerance. This is the same for LoRa-PHY. For example, the standard LoRa encoder uses a chirp symbol with $\frac{2^{\text{SF}}}{BW}$~s period to represent SF bits, thus the data rate is $\frac{\text{SF} \times BW}{2^{\text{SF}}}$. When the period is extended by setting a larger SF, the data rate is lowered, and a higher energy peak will be obtained by the dechirp to tolerate a lower SNR.
Instead of extending the period of a symbol beyond SF-12 that is unavailable on COTS LoRa nodes, we reduce the data bits represented by a symbol to lower the data rate while keeping the symbol period unchanged. 
Our SF-configuration based encoder exploits the diversity of different SF chirps to encode data. Specifically, given different SF configurations, the frequency of a base up-chirp increases from $-\frac{BW}{2}$ to $\frac{BW}{2}$ in different speed, and its period is $\frac{2^{\text{SF}}}{BW}$. 
We use a LoRa node to continuously transmit base up-chirps in SF configurations 9 to 12 in a clean LoS environment.
With the collected signals, we plot the amplitude and phase spectrograms derived by Short-time Fourier Transform (STFT) to illustrate the time-domain difference among the four SF configurations in Figure~\ref{fig:spectrogram_illustration}. 
As shown in Figure~\ref{fig:spectrogram_illustration}(d), given 125~kHz $BW$, an SF-12 base up-chirp occupies channel for 32.768~ms, which equals to the period of eight SF-9, four SF-10, or two SF-11 base up-chirps shown in Figure~\ref{fig:spectrogram_illustration}(a), (b), and (c).
In the settings, a \ours symbol encodes 2-bit data and has the same period as an SF-12 chirp. The data rate is lowered to $\frac{2BW}{2^{12}}$ which is similar to SF-15 chirps $\frac{1.875BW}{2^{12}}$. We prove that the encoder can be implemented on COTS LoRa nodes (\S\ref{sec-design-1}).

\vspace{1mm}
\noindent
\textbf{Encoder-Decoder Co-Design:} With the standard LoRa encoder, NELoRa~\cite{li2021nelora} has demonstrated that a DNN-based decoder can capture multi-dimensional features from both amplitude and phase spectrograms to lower the requested SNR in decoding compared with the dechirp. However, for the standard LoRa encoder with SF-12, the classification task of NELoRa is a 4,096-class (e.g., $2^{12}$ bit strings) classification problem. The size of the input SF-12 spectrograms is 15$\times$ larger than SF-7. The SNR gain is significantly degraded by exhausting computing.
%
%
%
With our SF-configuration based encoder, we adopt a neural-enhanced decoder~\cite{li2021nelora} to classify the SF configuration of a symbol. The benefits are two folds. First, the decoding problem becomes to a 4-class classification problem. Second, as shown in Figure~\ref{fig:spectrogram_illustration}, we can observe the different time-domain periodical patterns in amplitude and phase spectrograms, enabling a new feature space to distinguish different symbols. Thus, an extra SNR gain can be expected compared to the traditional SF-12 NELoRa decoder. We customize the input size and our DNN model design to reduce the computation overhead significantly while obtaining the same SNR gain (\S\ref{sec-design-3}).



\section{System Design}
\label{sec-system-design}


%

\subsection{SF-configuration-based Encoder}
\label{sec-design-1}

To be compatible with existing LoRa deployments, the encoder of \ours should conform to the LoRa standard that cannot arbitrarily alter the signal. 
We design an SF-configuration-based encoder by taking full advantage of the capabilities of COTS LoRa nodes without needing any hardware modification. Thus, \ours can be deployed in any operational LoRa system with minimum deployment cost.

\vspace{1mm}
\noindent
\textbf{Data modulation:}
COTS LoRa nodes use six SF configurations from 7 to 12 to provide different data rates for satisfying varying SNR situations.
Our encoder leverages SF combinations from all available SF configurations of COTS LoRa nodes for data encoding.
It uses four SF configurations to represent each 2-bit data, which provides three data rate settings, namely [7-10], [8-11], and [9-12], to tradeoff the SNR tolerance with the data rate.
In each setting, the period of an SF-configuration-based symbol is the period of an SF$_{\text{max}}$ chirp (i.e., SF-10, SF-11, and SF-12), as the example symbols shown in Figure~\ref{fig:cover} and Figure~\ref{fig:spectrogram_illustration}.

\begin{figure}[!t]
    \centering
    \includegraphics[width=1\linewidth]{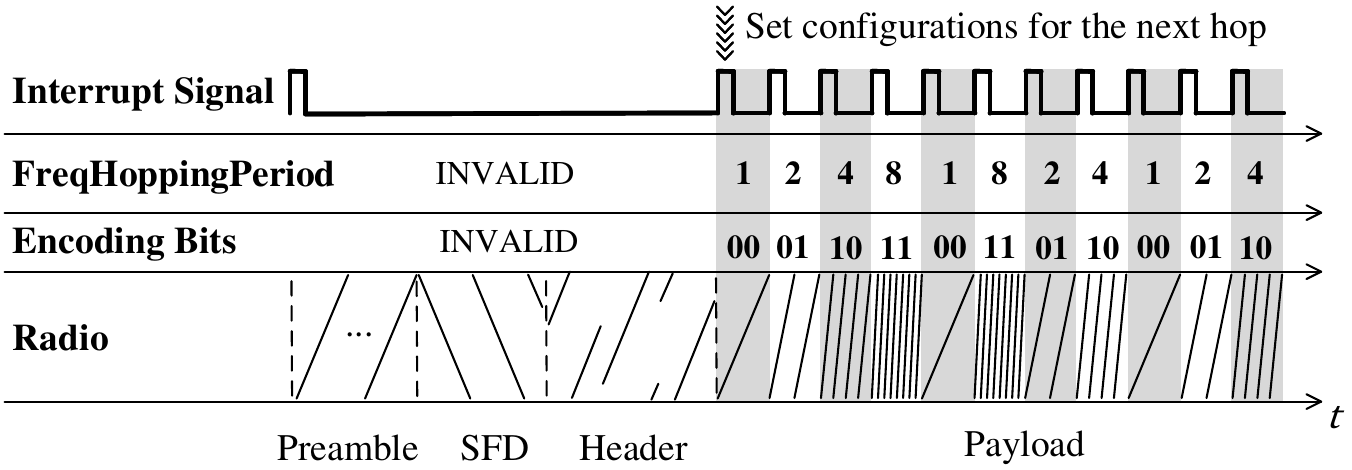}
    \caption{Timeline of the SF-configuration-based encoder. The encoder leverages hardware interrupts of COTS LoRa nodes for manipulating radio configurations and encoding data bits.}
    \label{fig:encoder_timeline}
\end{figure}

\vspace{1mm}
\noindent
\textbf{Frequency hopping enabled \ours packet:}
While \ours encoder changes the SF among different symbols of a single \ours packet, implementing such an encoder on COTS LoRa nodes offers radically new challenges.
LoRa-PHY uses a single SF configuration to transmit a LoRa packet. Before generating a packet, it sets all the modulation configurations, including carrier frequency, bandwidth, and SF.
We settle this challenge and implement our encoder based on the frequency hopping capability of COTS LoRa nodes~\cite{SX127X}.
The LoRa standard requires LoRa nodes to support frequency hopping, enabling long-duration packet transmission without violating the maximum permissible channel dwell time.
The key principle behind the frequency hopping scheme is hardware interrupt, named \textsf{ChangeChannelFhss}, that enables LoRa nodes to select and switch to a new channel during packet transmission.
After a predetermined hopping period, the transmitter and receiver change to the next channel in a predefined list of hopping frequencies to continue transmission and reception of the next portion of the packet.
Our key observation is that a LoRa node can modify not just the channel but all configurations every time it triggers the \textsf{ChangeChannelFhss} interrupt.
Therefore, we can implement an SF-configuration-based encoder on a COTS LoRa node without adding extra hardware by making the node periodically trigger the \textsf{ChangeChannelFhss} interrupt and change its SF configuration during a \ours packet transmission. Moreover, we keep the transmission on a single channel by setting the list of hopping frequencies consists items with identical frequency. 

Figure~\ref{fig:encoder_timeline} shows the timeline of the SF-configuration-based encoder for transmitting a \ours packet. The transmission starts with a standard LoRa-PHY preamble, start frame delimiter (SFD), and LoRa-PHY header, all at $SF_{max}$ as defined in the data modulation. 
At the beginning of each transmission, an interrupt signal \textsf{ChangeChannelFhss} is generated, where the interrupt handler programs the frequency, SF, and hopping period for the first hop of the payload.
The interrupt signal is cleared after all configurations have been settled.
Then, during payload transmission, the transmitter periodically triggers the \textsf{ChangeChannelFhss} interrupt to modify the SF configuration for \ours data modulation.
The time that each hop of transmission will dwell is determined by \textsf{FreqHoppingPeriod} which is an integer multiple of symbol periods.
As is illustrated in data modulation, the periods of all SF-configuration-based symbols should be identical to the period of an $SF_{max}$ chirp. Thus, we determine the \textsf{FreqHoppingPeriod} based on the SF configuration of chirps in the corresponding hop, i.e., 
$FreqHoppingPeriod = 2^{SF_{max} - SF}$
\blue{The new configurations are programmed within the current hopping period to ensure it has been set when the next hop begins. The interrupt computation is much shorter than the symbol on-air time, causing no extra latency.}
\begin{figure}[!t]
    \centering
    \includegraphics[width=1\linewidth]{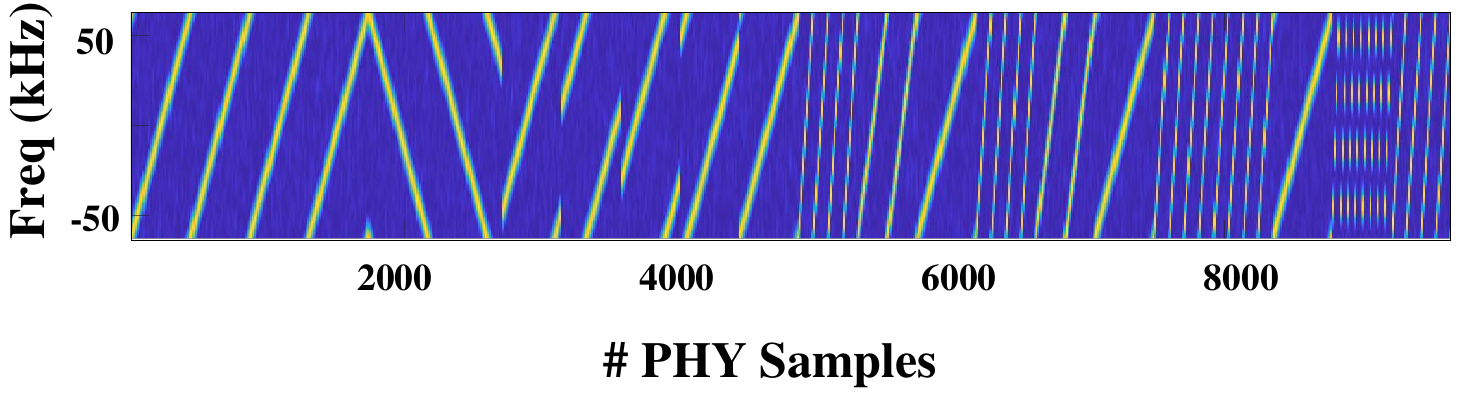}
    \vspace{-3mm}
    \caption{Spectrum of a \ours packet generated by the SF-configuration-based encoder.}
    \label{fig:frame_structure}
    \vspace{-3mm}
\end{figure}
We encode both the header and payload of \ours on the payload transmission of the standard LoRa by modifying SFs of symbols.
Therefore, we can implement \ours encoder with only COTS LoRa nodes.
Figure~\ref{fig:frame_structure} shows the spectrum of a real \ours packet generated by a COTS LoRa node with the SF-configuration-based encoder and recorded by a software-defined radio.


\begin{figure}[!t]
    \centering
    \includegraphics[width=0.45\textwidth]{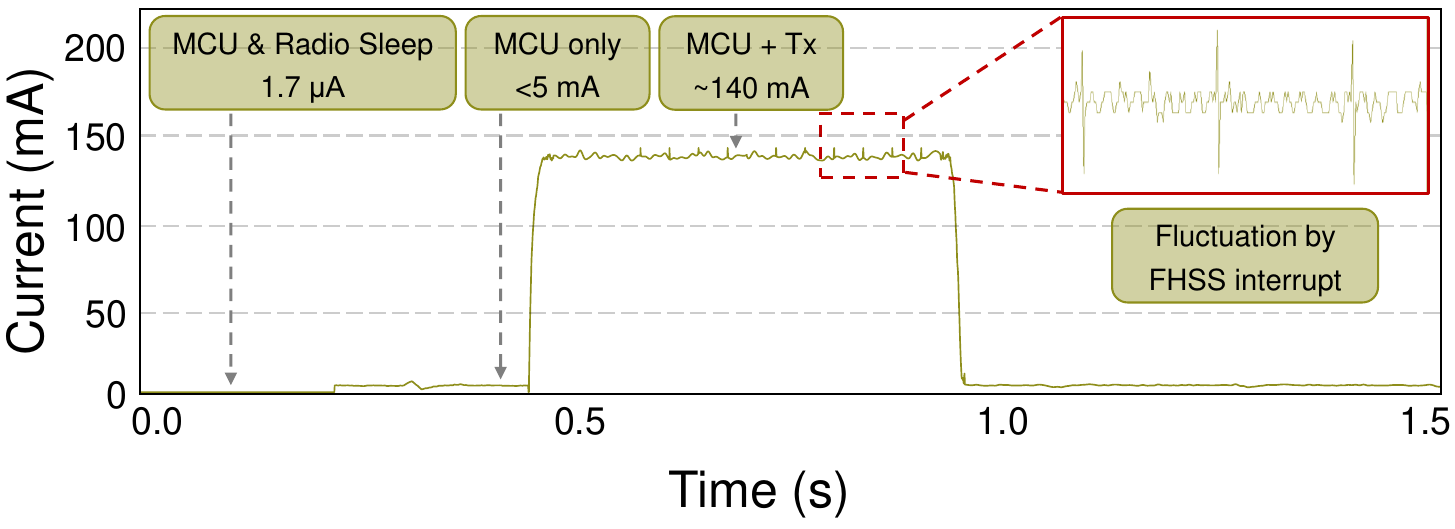}
    \vspace{-2mm}
    \caption{\blue{Current profile of symbol hopping in \ours.}}
    \label{fig:power_profile}
    \vspace{-3mm}
\end{figure}

\blue{Figure~\ref{fig:power_profile} shows the current profile of the transmission, including the instant current for all radio access phases. The LoRa node stays in sleep mode when it does not transmit data. The radio transmission consumes the highest amount of energy by a large margin. The power consumption of the MCU is much less than that of the radio circuit. 
Thus, the ChangeChannelFhss interrupt processed by the MCU only introduces a small current fluctuation during the packet transmission, which has a neglectable impact on the whole energy consumption of the LoRa node.}

\subsection{Neural-enhanced Decoder Customization}
\label{sec-design-3}

Inspired by the neural-enhanced decoder in NELoRa~\cite{li2021nelora}, \ours also aims to decode the SF-configuration based symbols in a neural-enhanced manner.
Observing the encoder difference between NELoRa and \ours, we customize \ours's neural-enhanced decoder in three folds, including the fixed-size input feature map, the customized DNN structure, and the phase-based data augmentation.

\vspace{1mm}
\noindent
\textbf{Feature Extraction:} 
Like NELoRa, our neural-enhanced decoder converts symbols into time-frequency spectrograms as the input feature maps. 
We first apply STFT on a symbol with the pre-configured sliding window size and frequency bin size. 
Then, we concatenate the real and imaginary parts to retain the amplitude and phase information of the symbol.
Compared with the dual-channel feature map of NELoRa~\cite{li2021nelora}, \ours tailors its input given its new encoding mechanism. 
Specifically, NELoRa~\cite{li2021nelora} requires the frequency resolution of the input feature maps larger than the number of output codes (i.e., $2^{SF}$).
However, \ours only aims for its configuration pattern extraction (i.e., 4-class classification) from the input feature maps. 
To derive the optimal input feature maps for prohibiting the noise and maximizing extra SNR gains, we evaluate the impact of the spectrogram resolution, which is determined by the number of frequency bins in STFT and the time-domain samples in each segment. 

\begin{figure}[!t]
    \centering
    \includegraphics[width=0.35\textwidth]{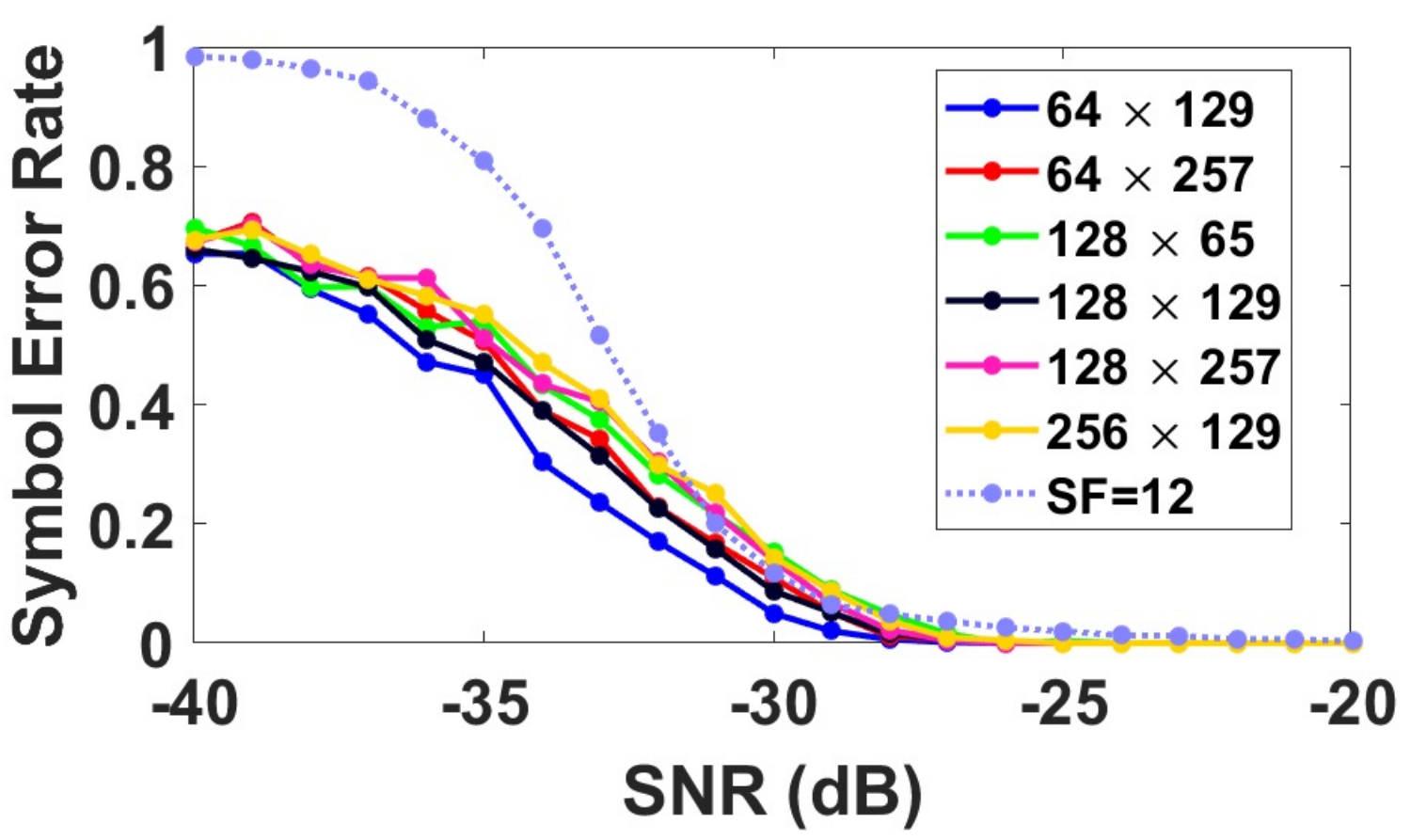}
    \caption{The impact of spectrogram resolutions (frequency bins $\times$ time-domain samples) on SNR gains.}
    \label{fig:dnn-stft}
    \vspace{-3mm}
\end{figure}

By testing on 2,000 \ours symbols with encoding setting of SF configurations SF-[9,12], we illustrate the SER performance for some input feature maps with different resolutions in Figure~\ref{fig:dnn-stft}. 
In comparison with the standard LoRa-PHY demodulation (i.e., SF-12, dashed purple line), a larger resolution (e.g.,256$\times$129, solid yellow line) of the spectrogram adversely reduces the SNR gains from our DNN decoder. 
In contrast, a coarse-grained input (e.g.,64$\times$129, solid blue line) achieves the largest SNR gains. 
Given the strong noise level at extremely-low SNRs, we argue that a coarse-grained input spectrogram can increase the SNR gains of our neural-enhanced decoder, making it more resilient to different noise patterns in the wild.

\begin{figure}[t!]
\centering
\includegraphics[width=0.9\linewidth]{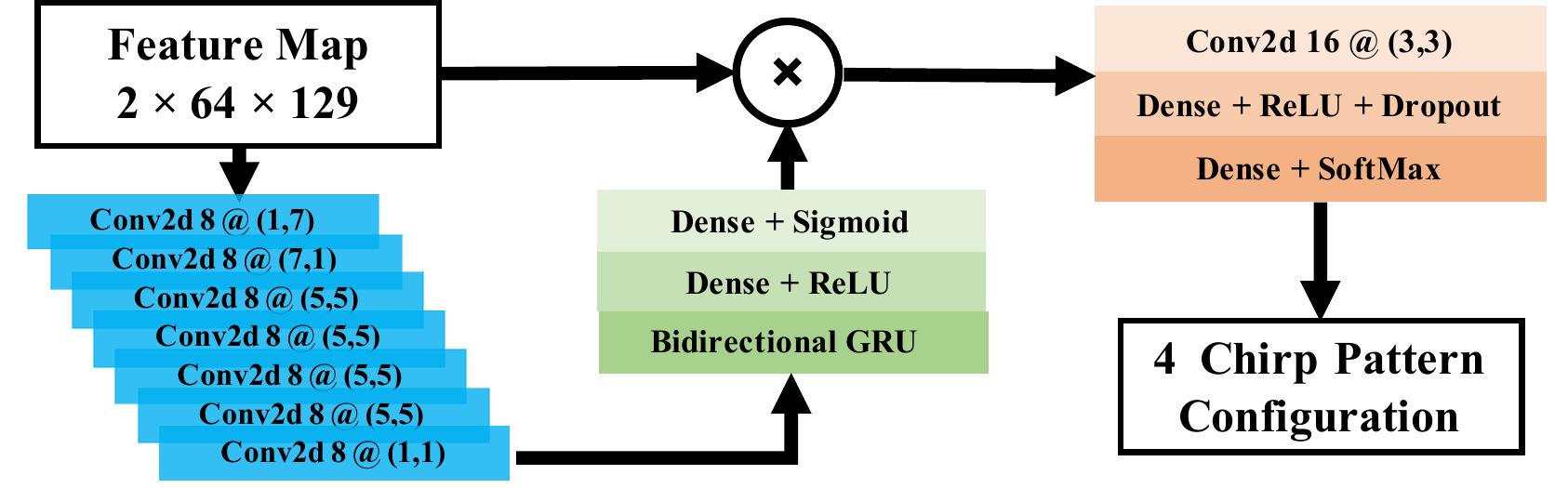}
\caption{The neural network structure of the neural-enhanced decoder for efficient training. We note the channel number and kernel size for convolutional layers.}
\label{fig:network}
\vspace{-3mm}
\end{figure}

\vspace{1mm}
\noindent
\textbf{DNN Model Customization:} Given the 4-class SF-based configuration recognition task, \ours can achieve a lightweight neural-enhanced decoder for efficient running. 
In other words, the DNN structure proposed by NELoRa~\cite{li2021nelora} is sub-optimal for \ours,
since NELoRa~\cite{li2021nelora} aims to extract the SF types of encoded initial frequency offsets with the standard LoRa encoding mechanism, which is much larger than our 4-class classification task. 

In comparison with NELoRa, we customize our neural-enhanced decoder in two folds:
First, for efficient training, we focus on optimizing the first four modules of NELoRa for mask generation and keeping the classification network unchanged for robustness.
Specifically, \ours simplifies the overall architecture of the neural-enhanced decoder by using a lightweight first Conv2d module with fewer filters and replacing the LSTM layer with a bidirectional GRU layer, which is more computation-efficient for temporal feature extraction.
Second, for inference running efficiency, we adopt the magnitude-based pruning method ~\cite{molchanov2019importance} to identify and prune redundant weights in dense layers in order to reduce the decoding time on edge devices while keeping high decoding reliability.
Specifically, we calculate the L1-norm of the weights in every dense layer and preserve those with the largest L1-norms.
Figure~\ref{fig:network} shows the concise network structure of \ours, which consists of seven modules in total. Specifically, the first four modules aim to generate a filter mask to be multiplied with the input spectrogram. Then \ours feeds the masked spectrogram into a three-module classifier for SF-based configuration recognition.

\begin{figure}[!t]
    \centering
    \subfloat[Per-layer analysis for an adaptive NELoRa based DNN model]{
    \label{fig:perlayer-1}
    \includegraphics[width=0.48\textwidth]{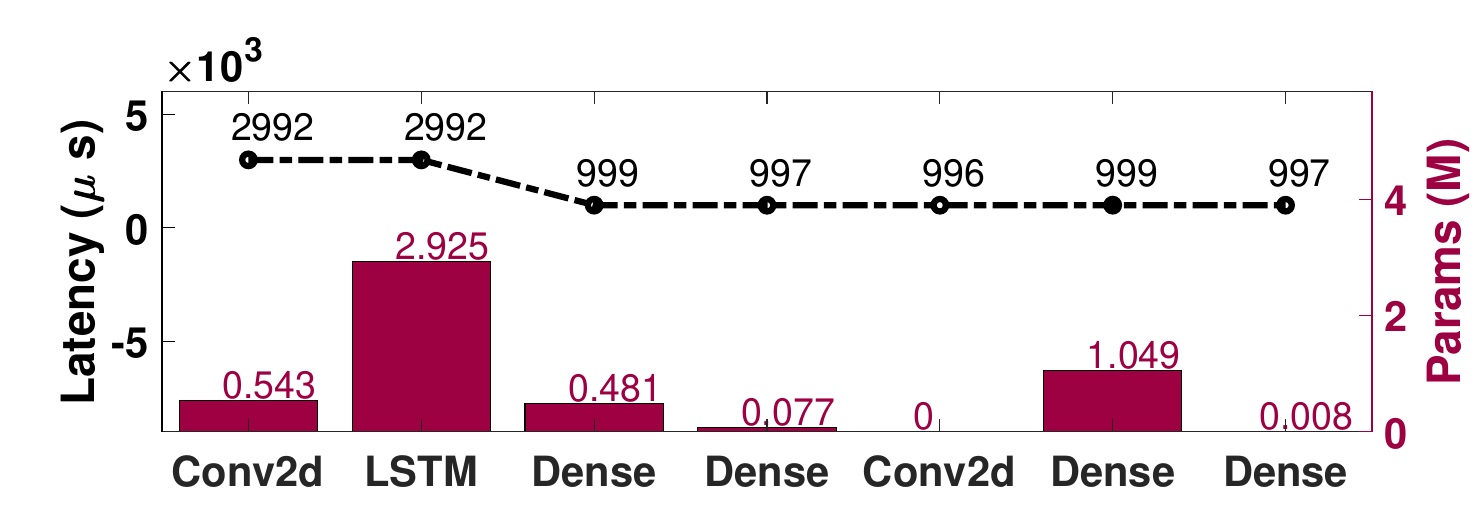}
    }\\
     \vspace{-4mm}
    \subfloat[Per-layer analysis for \ours]{
    \label{fig:perlayer-2}
    \includegraphics[width=0.48\textwidth]{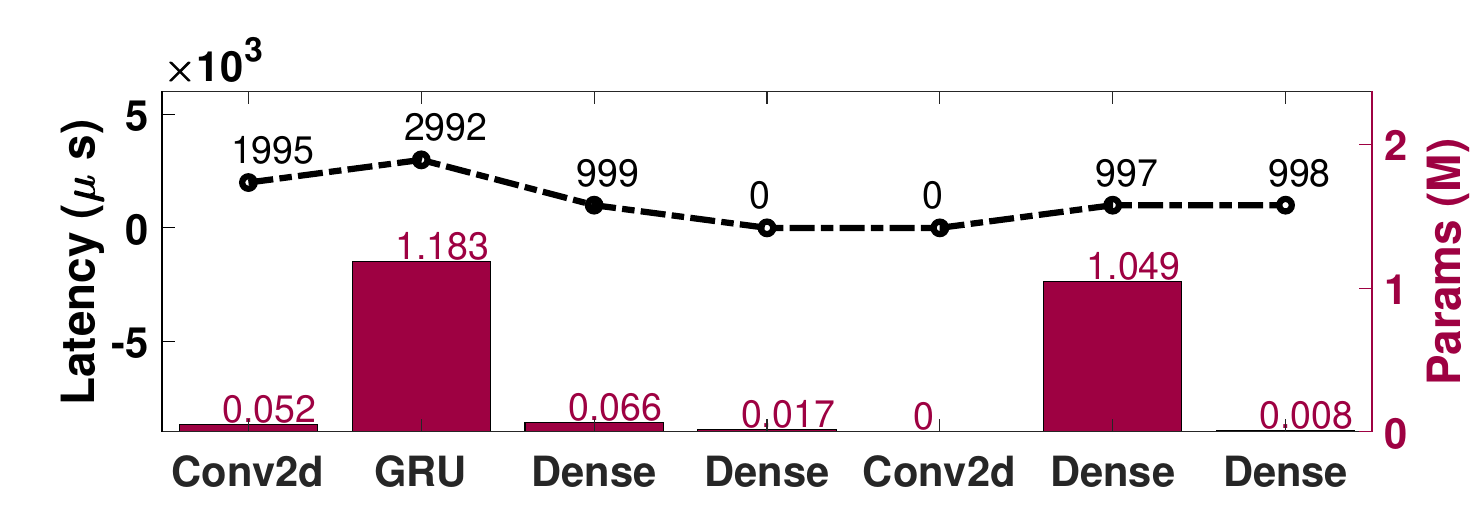}
    }
    \caption{Per-layer analysis of adaptive NELoRa (top) and \ours (bottom) for running efficiency over a packet decoding (i.e., 16 \ours symbols) on a PC with GPU 1050ti.}
    \label{fig:perlayer}
\vspace{-6mm}
\end{figure}

We empirically study the efficiency of our simplified neural-enhanced decoder against an adaptive NELoRa~\cite{li2021nelora} based DNN decoder. Specifically, for a fair comparison, we adapt the input and output of NELoRa's DNN decoder to our 4-class classification problem and compare the adaptive DNN model with \ours's in the same settings.
Figure~\ref{fig:perlayer} shows the per-layer analysis on running time and parameter size~\cite{flops_counter}.
By testing on 50 \ours packets (i.e., 16 symbols) with an encoding setting of SF-[9,12], the average total running time is around 11~ms to decode a packet for the adaptive NELoRa~\cite{li2021nelora} on a local PC equipped with a GPU 1050ti. \ours achieves 8~ms for its packet decoding.
Besides, the parameter size of \ours is 2.3~M, less than half of the adaptive NELoRa's 5.3~M parameters.

\vspace{1mm}
\noindent
\textbf{Few-shot Learning:} 
In machine learning, few-shot learning~\cite{sung2018learning} refers to problems with a limited training dataset but make a prediction about many unknown examples in the future.
In \ours, the phase of a \ours symbol is a useful feature space captured by the DNN model. 
According to the process of symbol phase calibration in $\S$~\ref{sec-design-2}, a limited training dataset can only provide \ours symbols with randomly scatted initial phases induced by carrier frequency offset (CFO) and sampling frequency offset (SFO), making it hard for the neural-enhanced decoder to recognize the chirps with unseen phase offset patterns.
During training, we explore the potential feature space of unseen chirp symbols and their phase offset patterns as much as possible to adapt to the various scenarios in the wild.
Based on our empirical study on initial phase distribution, we adopt the data augmentation~\cite{wang2020push} to generate synthetic symbols in the training stage, in which we augment the initial phase manually, enabling its uniform distribution in the training dataset. Thus, we can use data augmentation to reduce the model retraining overhead to achieve an ``one-fits-all'' model training process.

\begin{figure}[t]
    \centering
    \includegraphics[width=0.4\textwidth]{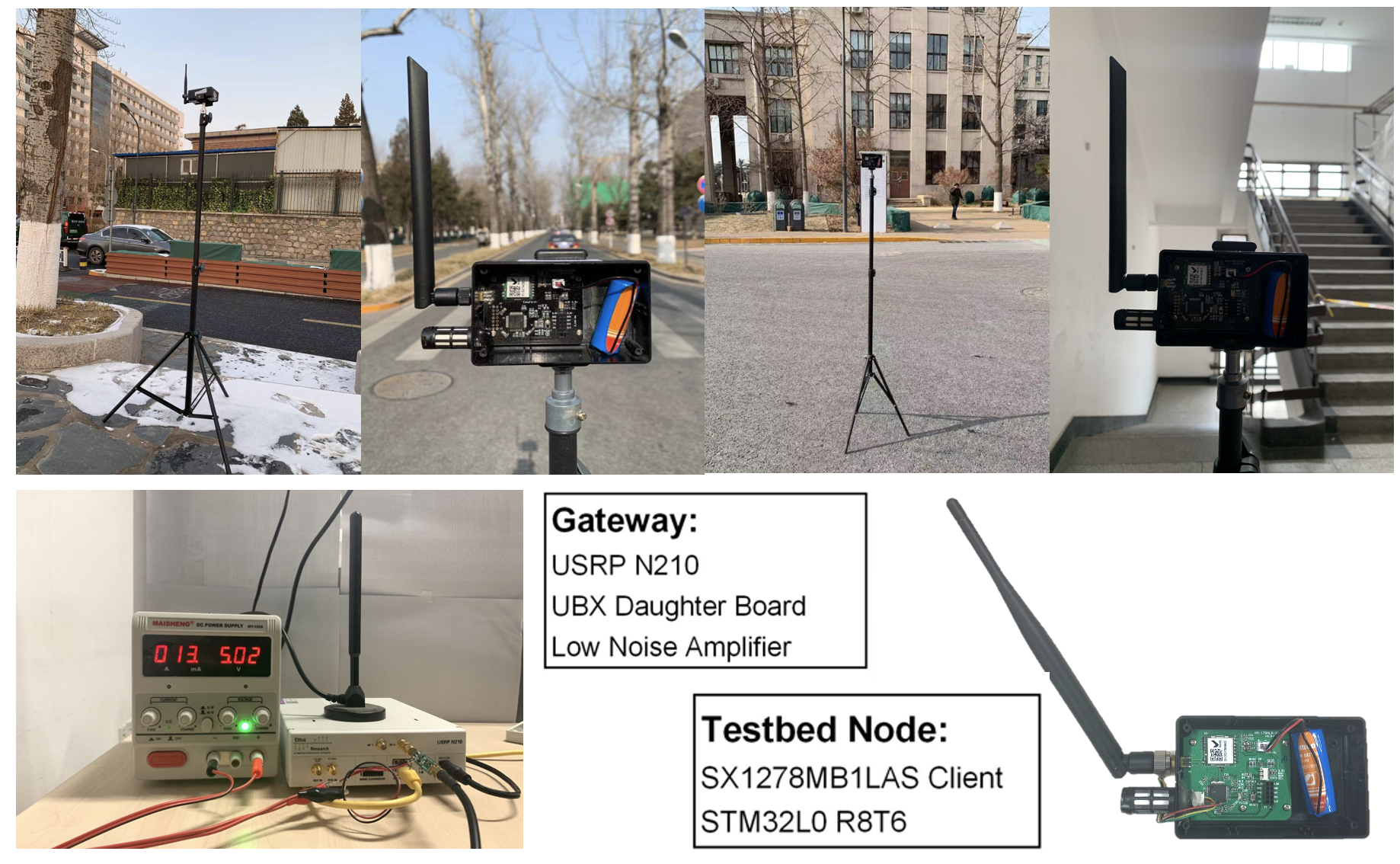}
    \caption{USRP N210 based gateway and COTS LoRa nodes deployed in a building and a university campus.}
    \label{fig:software-hardware}
\vspace{-3mm}
\end{figure}

\subsection{Packet Detection}
\label{sec-design-2}

After a LoRa node transmits a \ours packet through the air, the packet should be reliably detected at the gateway side even with extremely low SNR. Then, its payload is divided into multiple symbols and fed into the DNN-based decoder for SF configuration recognition.
Given a period of received signals, we identify whether they contain a LoRa-PHY preamble, which consists of a series of $\text{SF}_{\text{max}}$ base up-chirps, as an indicator of whether a \ours packet is coming. 
%
We divide the received signals into $N$ symbol-length signal segments, where $N$ is the number of base up-chirp symbols in a LoRa-PHY preamble.
Then, we combine the $N$ segments as a superposed signal segment. All the base up-chirp symbols are constructively superposed if a preamble is in the received signals~\cite{li2021nelora}.
%

Then, we use a standard base up-chirp symbol to calculate its cross-correlation with the superposed signal segment. We treat the detection of a significant correlation peak as a sign of a successful preamble detection. The threshold for the peak detection is based on the channel estimation, which is $6$ standard deviations of the mean noise correlation.
The index of the correlation peak also indicates the boundary of the received LoRa preamble symbols. Therefore, we can align the timing of the demodulation window and extract the aligned \ours symbols from the payload part based on the detected preamble.

\section{Implementation}
\label{sec_implementation}

We have implemented \ours on a Software Defined Radio (SDR) and COTS LoRa nodes, shown at the bottom of Figure~\ref{fig:software-hardware}. Specifically, the USRP N210 SDR platform can capture over-the-air LoRa signals by operating on a UBX daughter board with a sampling rate of 1~MS/s. 
The captured signal samples are then delivered to a back-end host, preprocessed in MATLAB, and demodulated by the neural-enhanced decoder in Pytorch.
The COTS SX1278 based LoRa nodes transmit \ours packets with random payloads applying the SF-configuration-based encoding mechanism periodically. They are deployed in both indoor (a building) and outdoor environments (campus), as shown at the top of Figure~\ref{fig:software-hardware}. 

\subsection{Baseline Methods}
\label{subsec-baseline-methods}

Besides LoRa-PHY, we adopt/develop three baseline encoder-decoder co-design methods to fairly understand the performance of \ours in terms of SNR threshold at extremely-low SNR conditions. 
All these methods are implemented and evaluated with the same hardware and environments of \ours.

\begin{figure}
\centering
\includegraphics[width=0.75\linewidth]{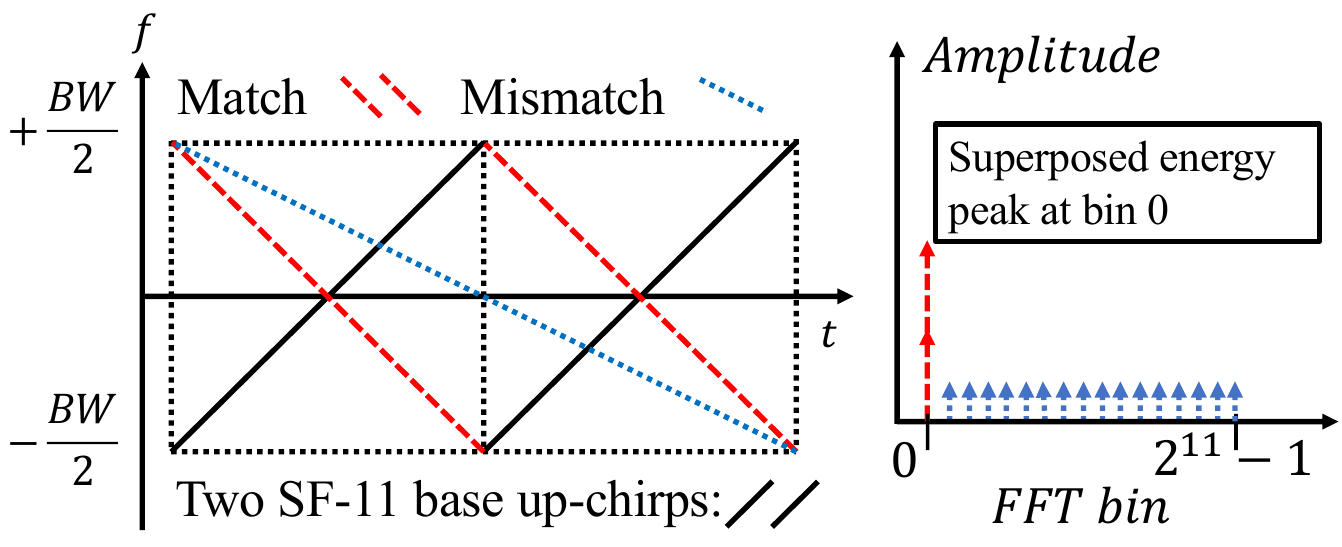}
\vspace{-3mm}
\caption{After multiplying the matched base down-chirp template, a superposed energy peak appears at bin 0.}
\label{fig:SNR-gain-reason}
\end{figure}

\blue{
\noindent
\textbf{Cor: \ours encoder + correlation-based decoder.}
We use chirp coherent combining ($\S$~\ref{sec-design-2}) to develop a correlation-based spectrum-energy decoder sitting on our SF-configuration based encoder. Figure~\ref{fig:SNR-gain-reason} illustrates an SH-[9,12] symbol with two SF-11 base up-chirps, we multiply two matched base down-chirps (called \emph{template}) and coherently combine the derived signals in the two SF-11 windows together. After applying FFT, a superposed energy peak appears at bin 0. For using a mismatched template like an SF-12 base down-chirp, the energy is dispersed at all bins without any peak. Hence, we can search the expected energy peak from several templates with different SF configurations to find the matched one. \textbf{Cor} has three configurations with same encoder of \ours (SH[7-10], SH[8-11], SH[9-12]) but different decoding scheme.
}
%


\noindent
\textbf{Ostinato: repetitive SF-12 chirps based encoder + chirp coherent-combining based decoder.}
Ostinato~\cite{xu2022ostinato} can be regarded as a special case of \ours's chirp repeating encoding. Specifically, Ostinato uses repeated SF-12 chirps, having the identical initial frequency offset, to encode the same data as a single SF-12 chirp does. Then, the decoder coherently combines the multiple SF-12 chirps to obtain a more obvious energy peak during dechirp than any individual SF-12 chirp. We implement the encoder by using our chirp repeating feature to generate the repeated SF-12 chirps and adopt the phase calibration method to fine-tune the chirp coherent combining for reliable decoding. It tree
configurations using two, four, and eight repeated SF-12 chirps to compose a symbol.

%
%

\begin{figure}[t!]
\centering
\includegraphics[width=1\linewidth]{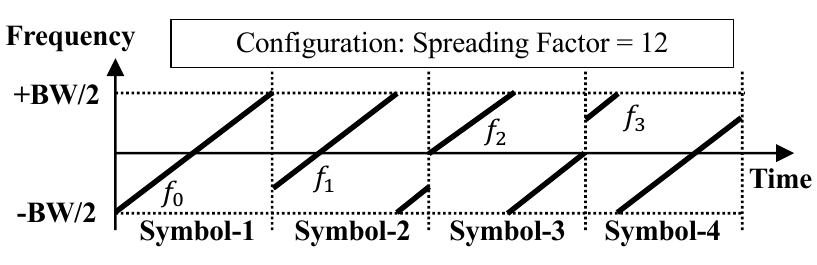}
\caption{The illustration of a neural-efficient LoRa encoder, using four initial frequency offsets (e.g., $f_0$, $f_1$, $f_2$, $f_3$) to represent 2-bit data.}
\vspace{-3mm}
\label{fig:encoder-2bit}
\end{figure}

\vspace{1mm}
\noindent
\textbf{IFO-2: Initial-frequency-offset-2-bit encoder + neural-enhanced decoder.}
With the similar design principle of our SF-configuration based encoder, we use four chirp symbols with different initial frequency offsets and the same SF configuration to encode 2-bit data.
Specifically, as shown in Figure~\ref{fig:encoder-2bit}, we select four SF-12 chirps with deterministic initial frequency offsets (e.g., $f_0$, $f_1$, $f_2$, $f_3$) to encode 2-bit data.
%
%
We set the initial frequency offsets as $-\frac{BW}{2}$, $-\frac{BW}{4}$, $0$ and $\frac{BW}{4}$ to distinguish these symbols as much as possible in the feature space. To achieve similar training and computation overhead, we adopt the same input size and structure with \ours's DNN model to decode data in a neural-enhanced manner. In addition, we adopt the same few-shot learning method as \ours to train the neural-enhanced decoder.
To keep similar data rate to \ours, \textbf{IFO-2} has three configurations which use SF-10, SF-11, and SF-12 chirps to form symbols corresponding to SF-[7-10], SF-[8-11], and SF-[9-12] symbols in \ours. The decoding problem becomes classifying the initial frequency offsets, which is still a four-class classification. We adopt the same deep neural network (DNN) structure in Figure~\ref{fig:network} to classify symbols for decoding.

\section{Evaluation}
\label{sec-evaluation}

In this section, we evaluate the performance of \ours to answer the following questions.

\noindent    
$\bullet$ \textbf{Q1 (\S6.1)}: \textit{Compared to various baseline methods, does \ours achieve reliable LoRa communication with the maximum SRN gain beyond the LoRa-PHY SF-12 chirps?}

\noindent    
$\bullet$ \textbf{Q2 (\S6.2)}: \textit{Compare to NELoRa~\cite{li2021nelora}, does \ours reduce the system and computation overhead of the neural-enhanced decoder?}

\noindent
$\bullet$ \textbf{Q3 (\S6.3)}: \textit{Compared to LoRa-PHY with the same chirp period, does \ours enable reliable LoRa transmissions at extremely-low SNR in an indoor NLoS environment?}

\noindent    
$\bullet$ \textbf{Q4 (\S6.4)}: \textit{Is a pre-trained \ours general enough to be deployed in new scenarios for quick deployment?}

\noindent    
$\bullet$ \textbf{Q5 (\S6.5)}: \textit{Does \ours perform consistently at the outdoor campus-scale testbed with complex NLoS scenarios and various noise patterns?}

\vspace{1mm}
\noindent
\textbf{Metrics:} We utilize two main metrics for the experiments on our indoor and outdoor testbeds. \\
$\bullet$ \textit{Symbol Error Rate (SER)} is widely used to demonstrate the channel noise resilience of a physical layer design given different SNR~\cite{tong_combating_2020,tong_colora_2020,li2021nelora}. A low SER is desirable. \\ 
$\bullet$ \textit{SNR Threshold} is the lowest SNR that a physical layer design can keep the SER under a predetermined value. By default, we set the predetermined SER as 1\%.
Since a LoRa-PHY packet consists of multiple chirp symbols, a non-zero SER increases the possibility of packet corruption exponentially. However, we can use sophisticated coding mechanisms to tolerate specific symbol errors, and both LoRa-PHY and \ours can benefit from the error correction for reliable packet delivery. Hence, setting SER as 1\% is fair to measure the SNR threshold but does not degrade the packet transmission reliability in practice.

\noindent
\textbf{Method Performance Comparison:} We compare \ours with the two baselines ($\S$~\ref{sec_implementation}), which are denoted as \textbf{}, \textbf{Ostinato} and \textbf{IFO-2}. \textbf{Ostinato} adopts our SF-configuration based encoder that has three configurations indicated as \textbf{Ostinato}-[7,10], \textbf{Ostinato}-[8,11], and \textbf{Ostinato}-[9,12]. 
\textbf{IFO-2} configures its data rate with different SF chirps (e.g., 10, 11, 12) denoted as \textbf{IFO-2}-10, \textbf{IFO-2}-11, and \textbf{IFO-2}-12.
In addition, we compare \ours with LoRa-PHY, the standard LoRa physical layer, which uses initial frequency shift and the dechirp for encoding and decoding~\cite{eletreby_empowering_2017}. 
To achieve the highest SNR tolerance, we select the largest SF in the \ours's encoding setting as the benchmark of LoRa-PHY. For example, we adopt SF-10, SF-11, and SF-12 LoRa-PHY as benchmark for our encoding settings using SF-[7,10], SF-[8,11], and SF-[9,12], respectively. We denote LoRa-PHY as \textbf{Dechirp}.


\vspace{1mm}
\noindent
\textbf{Evaluation Dataset:} Our datasets consist of three parts.\\
$\bullet$ \textbf{Training and In-domain Testing:} 
We collect 20 symbols with original SNR higher than 30~dB for each \ours symbol configuration, 80\% of which are used to train the neural-enhanced decoder and the remaining 20\% for in-domain testing since both of them are from the same environment.
Note that we augment the training dataset by manually adding Gaussian white noises~\cite{tong_combating_2020,tong_colora_2020,li2021nelora} for fine-grained SNR control (e.g., [-50, 20]dB) and calibrating the initial phase of each symbol.
Thus, we can derive a ``one-fits-all'' neural-enhanced decoder by training only once, then apply it directly for the testing in a new environment, without data recollection or model re-training. \\
$\bullet$ \textbf{Cross-domain Testing:} To verify the robustness and generalization of our ``one-fits-all'' neural-enhanced decoder, we collect 20 new symbols with original SNR around 30~dB by setting three different LoRa nodes at three different environments. The different LoRa nodes result in different CFO and SFO patterns in the received signals. The three environments are three different office rooms with different spaces and furniture, resulting in different noise patterns. We further emulate the symbols at different SNRs (e.g., [-50, 20]dB) with the same SNR control method above. \\
$\bullet$ \textbf{A Campus-scale Testing:} To demonstrate \ours effeteness in the wild, we collect tens of packets with SF-[9-12] encoding configuration in different positions on our campus-scale testbed. In the outdoor testbed, buildings and trees introduce the NLoS communication scenarios. In different directions, the landcover diversity incurs different levels of signal path loss.


\blue{
\subsection{Communication Reliability Analysis}
\label{subsec-evaluation-2}

\noindent
\textbf{Setup:} We evaluate the SNR threshold of \ours under different symbol hopping encoder SH-[7-10], SH-[8-11], and SH-[9-12] configuration. 
We further evaluate the SNR threshold of \textbf{Cor}, \textbf{IFO-2} and \textbf{Ostinato} with all three encoding configurations. 

\begin{figure}[t]
    \centering
    \includegraphics[width=0.45\textwidth]{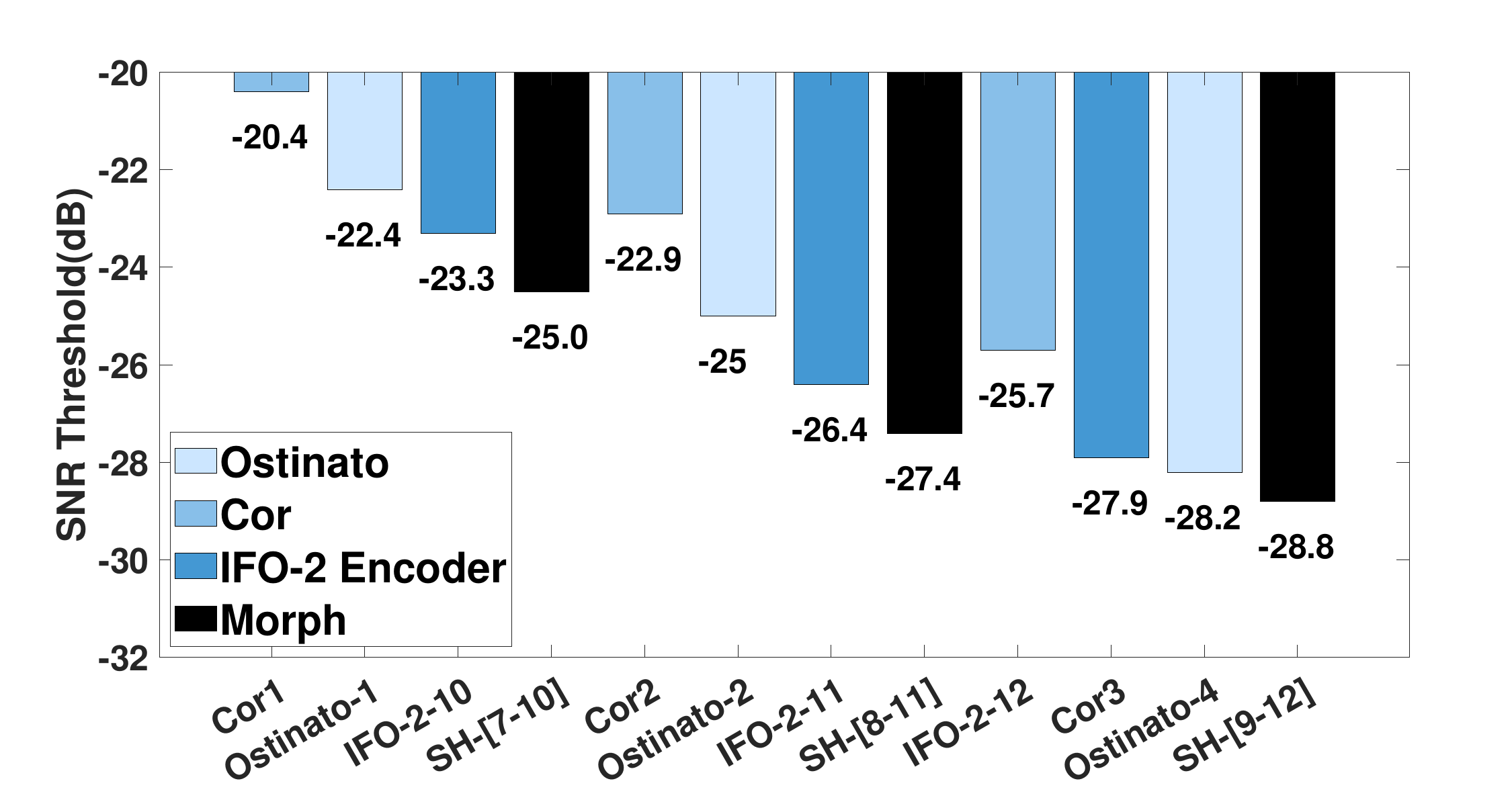}
    \caption{\blue{The comparison of SNR threshold and data rate among \ours, \textbf{Ostinato}, \textbf{IFO-2}, and Dechirp under various configurations.}}
    \label{fig:data_rate}
    \vspace{-4mm}
\end{figure}

\noindent
\textbf{Results:}
The results are shown in Figure~\ref{fig:data_rate}. \\
\noindent
\textbf{General SNR Gain:} The SNR thresholds of symbol hopping encoder are -24.5~dB, -27.4~dB and -28.8~dB using SH-[7-10], SH-[8-11], and SH-[9-12] encoding settings, separately. Compared with -22.4~dB SNR threshold of the standard LoRa under SF-12, we achieve the maximum 6.4~dB SNR gain, which is much larger than the approximate 2~dB SNR gain of NELoRa~\cite{li2021nelora}. This indicates that \ours significantly improves the communication reliability of LoRa. 
LoSee~\cite{losee22ren}, an urban LoRa measurement study, introduces a link model to predict the packet delivery ratio (PDR) within a square coverage area based on link SNR. The coverage area is defined where the overall PDR value exceeds $70\%$.  With the link model of LoSee and 6.4~dB SNR gain, \ours can achieve approximately $2.38\times$ coverage area compared with the standard LoRa in the urban environment.\\
\textbf{SNR Gain from the SF-configuration based encoder:} \textbf{IFO-2} adopts the same decoder with \ours, but encodes data with the same type of chirp signals. We can see that the SNR thresholds of \textbf{IFO-2} are 1.2~dB, 1.0~dB, and 0.9~dB higher than \ours with the identical settings of $SF_{max}-10$, $SF_{max}-11$, and $SF_{max}-12$. On average, \ours can tolerate 1.03~dB lower SNR than \textbf{IFO-2}. This verifies our SF-configuration based encoder provides a larger feature space than \textbf{IFO-2}'s 4-frequency-offsets based encoder, which makes the different \ours symbols can be easily distinguished with a DNN model. The SNR thresholds of SH-[7,10], SH-[8,11], and SH-[9-12] are 2.1~dB, 2.4~dB, and 0.6~dB lower than \textbf{Ostinato}-1, \textbf{Ostinato}-2, and \textbf{Ostinato}-4, separately.  On average, \ours achieves 1.7~dB SNR gain compared to \textbf{Ostinato}. This indicates our symbol hopping feature outperforms the chirp repeating pattern in \textbf{Ostinato} with complicated phase calibration for chirp coherent-combining.

\noindent
\textbf{SNR Gain from neural-enhanced decoder:} \ours and \textbf{Cor} utilize the same encoder but different decoders. \ours outperforms \textbf{Cor} 3.9~dB on average across all three encoding configurations. This verifies our neural-enhanced decoder can provide consistent SNR gains while the spectrum energy peak feature based \textbf{Cor} decoder suffers from inefficient signal coherent combining due to the phase offset. 

}

\begin{table}\footnotesize
\centering
\caption{System Overhead and Running Time Comparison}
\begin{tabular}{c|ccc}
\toprule
   & \textbf{\ours} & Ada. NELoRa & Comp. NELoRa$^{\mathrm{a}}$  \\
   \hline
Total Memory     & 17.19~Mb & 54.0~Mb & 152.9~Mb \\
Inference Time   & 0.26~s & 1.24~s & 0.97~s \\
\hline
\multicolumn{4}{l}{$^{\mathrm{a}}$ Compressed DNN model for SF-10 chirps}
\end{tabular}
\label{tab.sys_overhead}
\vspace{-4mm}
\end{table}

\subsection{System Overheads and Running Time}
\label{subsec-evaluation-3}

\noindent
\textbf{Setup:} we compare the system overhead and running time of \ours with an adaptive NELoRa based DNN model ($\S$~\ref{sec-design-3}) denoted as \textbf{Ada. NELoRa} and the compressed NELoRa~\cite{li2021nelora} denoted as \textbf{Comp. NELoRa}.
We use a Raspberry Pi to imitate the limited computation resource on a LoRa gateway to perform different models. We use Pytorch Profiler~\cite{pytorchprofiler} to measure the memory footprint and inference time of Ada. NELoRa and \ours on Rasberry Pi. The input size of Ada. NELoRa and \ours are fixed (e.g., 64$\times$129) for all three different configurations, and the model training and testing results are based on the in-domain dataset. We directly use the experimental results reported in NELoRa for Comp. NELoRa. To align the SNR tolerance of the Comp. NELoRa to \ours as much as possible, we adopt the compressed DNN model of NELoRa for SF-10 chirp symbols, which are the highest SF that NELoRa can support. We use two metrics as follows. One is total memory indicating the memory consumption of all layers.
The other is the inference time obtained by taking the average of 100 runs.

\noindent
\textbf{Results:} Table~\ref{tab.sys_overhead} lists the system overhead and running time of ours and others.
Using \ours, the inference time of a symbol is 0.26~s, and the total memory is 17.19~Mb.
Compared to Ada. NELoRa, \ours achieves $4.77\times$ inference speedup and $3.14\times$ total memory reduction.
This verifies the efficiency of our DNN model optimization and compression.
Compared to Comp. NELoRa, \ours achieves $3.73\times$ inference speedup and $8.89\times$ total memory reduction. The improvement will be getting larger when we train another Comp. NELoRa for decoding larger SF chirp symbols. The improvement verifies \ours is much more neural-efficient to fit the LoRa transmission at extremely-low SNR conditions.
In addition, such small inference latency and low system overhead allow us to decoder symbols in real-time, avoiding the several seconds of network latency and infrastructure cost when offloading the computation to the cloud.
Moreover, we notice that Comp. NELoRa's total memory is larger than Ada. NELoRa, but the relationship between their inference time is reversed. The reason is that Comp. NELoRa compresses some computing-intensive layers of NELoRa, which are still remained the same in Ada. NELoRa. Thus, the inference time compromises with the computation overhead of the computing-intensive layers in Ada. NELoRa. On the other hand, the input size of Ada. NELoRa (i.e., 64$\times$129) is much smaller than Comp. NELoRa (i.e., 1024$\times$33~\cite{li2021nelora}) leading total memory efficiency. 

\begin{figure}[!t]
    \centering
    \vspace{-3.5mm}
    \subfloat[In-domain SER Distribution]{
    \label{subfig-compression-ser}
    \includegraphics[width=0.25\textwidth]{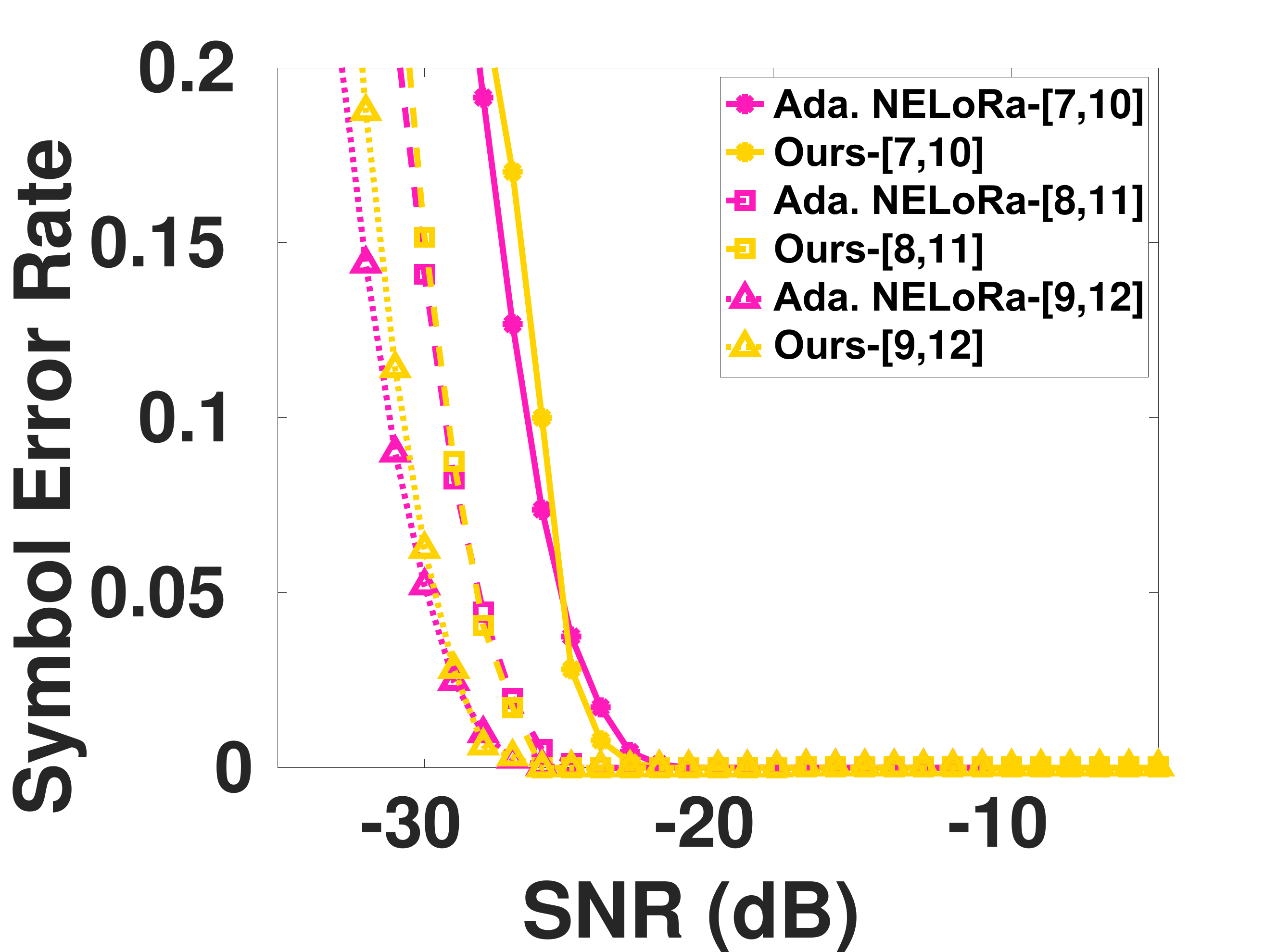}
    }
    \subfloat[In-domain SNR Threshold]{
    \label{subfig-compression-snr-gains}
    \includegraphics[width=0.24\textwidth]{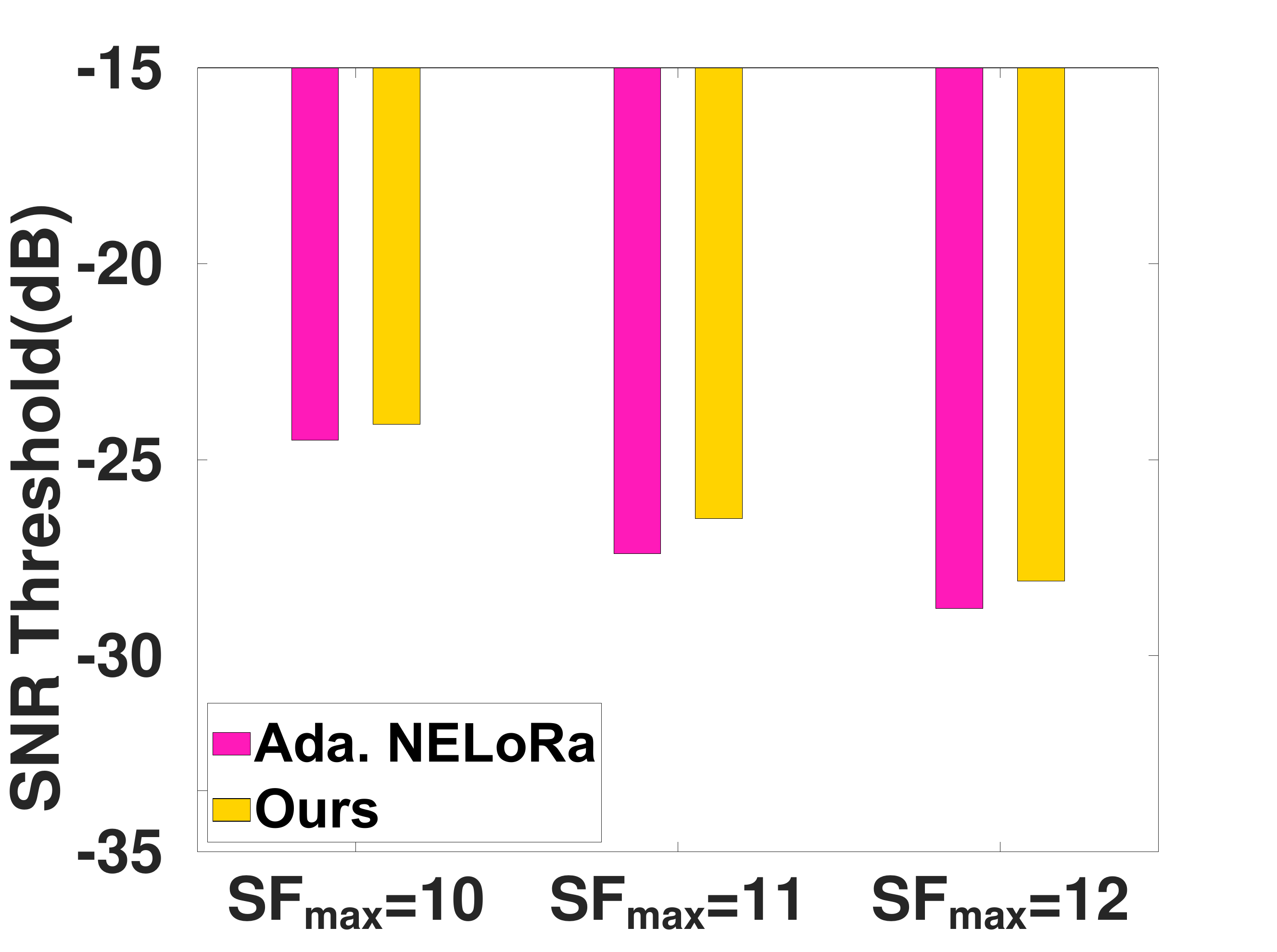}
    }
    \vspace{-2mm}
    \caption{Decoding performance comparison with or w/o \ours's DNN model customization.}
    \label{fig:compression}
    \vspace{-3mm}
\end{figure}

We also compare the \ours with Ada. NELoRa on the in-domain dataset.
Illustrated in Figure~\ref{subfig-compression-ser}, \ours demonstrates comparable SER in comparison with Ada. NELoRa under all three encoding setting SF-[7,10], SF-[8,11], and SF-[9,12]. In addition, the SNR thresholds of \ours are only 0.4-0.9~dB higher than Ada. NELoRa as shown in Figure~\ref{subfig-compression-snr-gains}. Overall, the model compression of \ours significantly improves the system efficiency while the resulting classification accuracy degradation brings only a minor SNR loss.

\begin{figure}[t]
    \centering
    \subfloat[In-domain SER Distribution]{\includegraphics[width=0.24\textwidth]{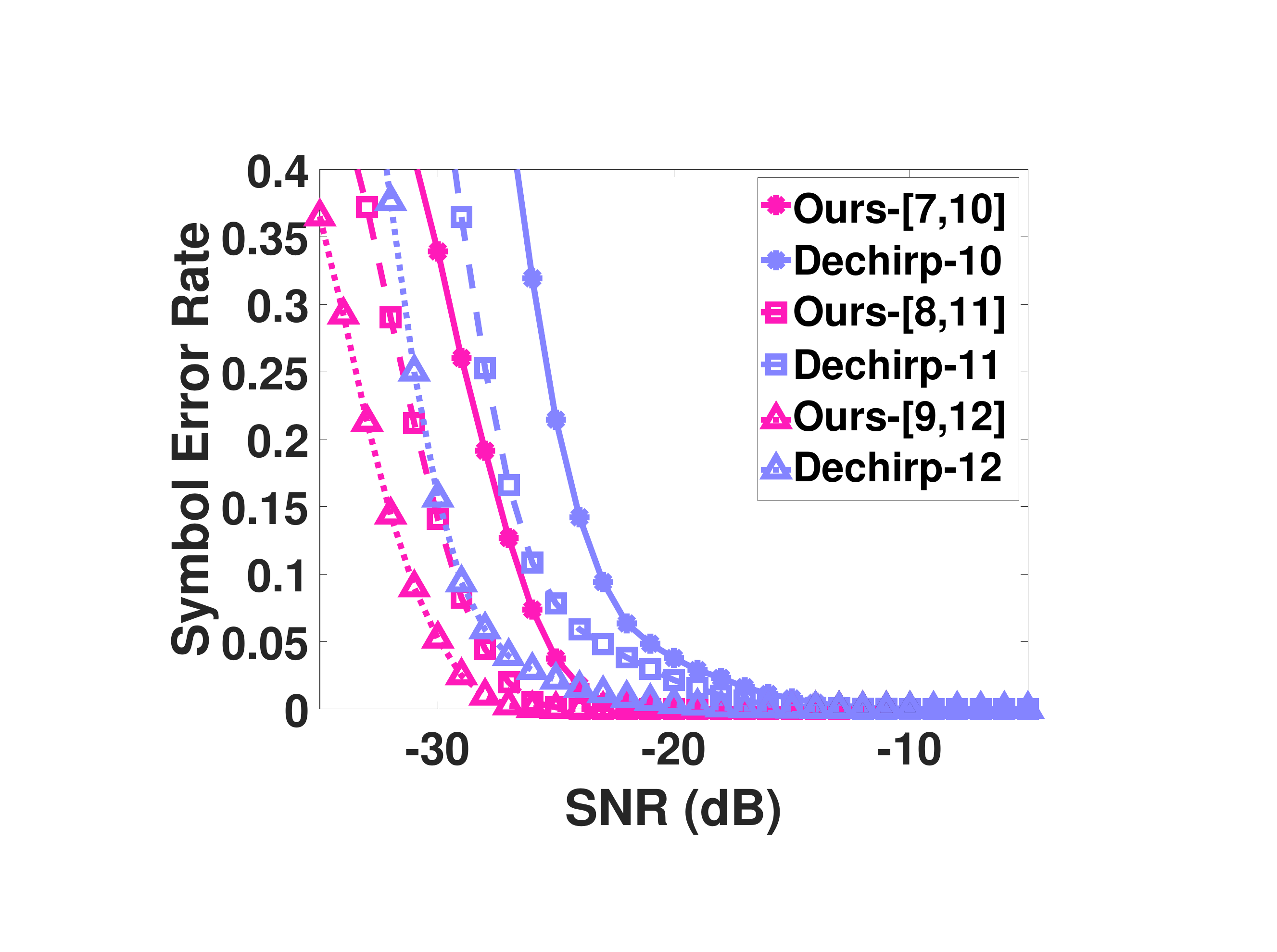}\label{fig:overall-performance-1}}
    \vspace{-3mm}
    \subfloat[In-domain SNR Threshold]{\includegraphics[width=0.24\textwidth]{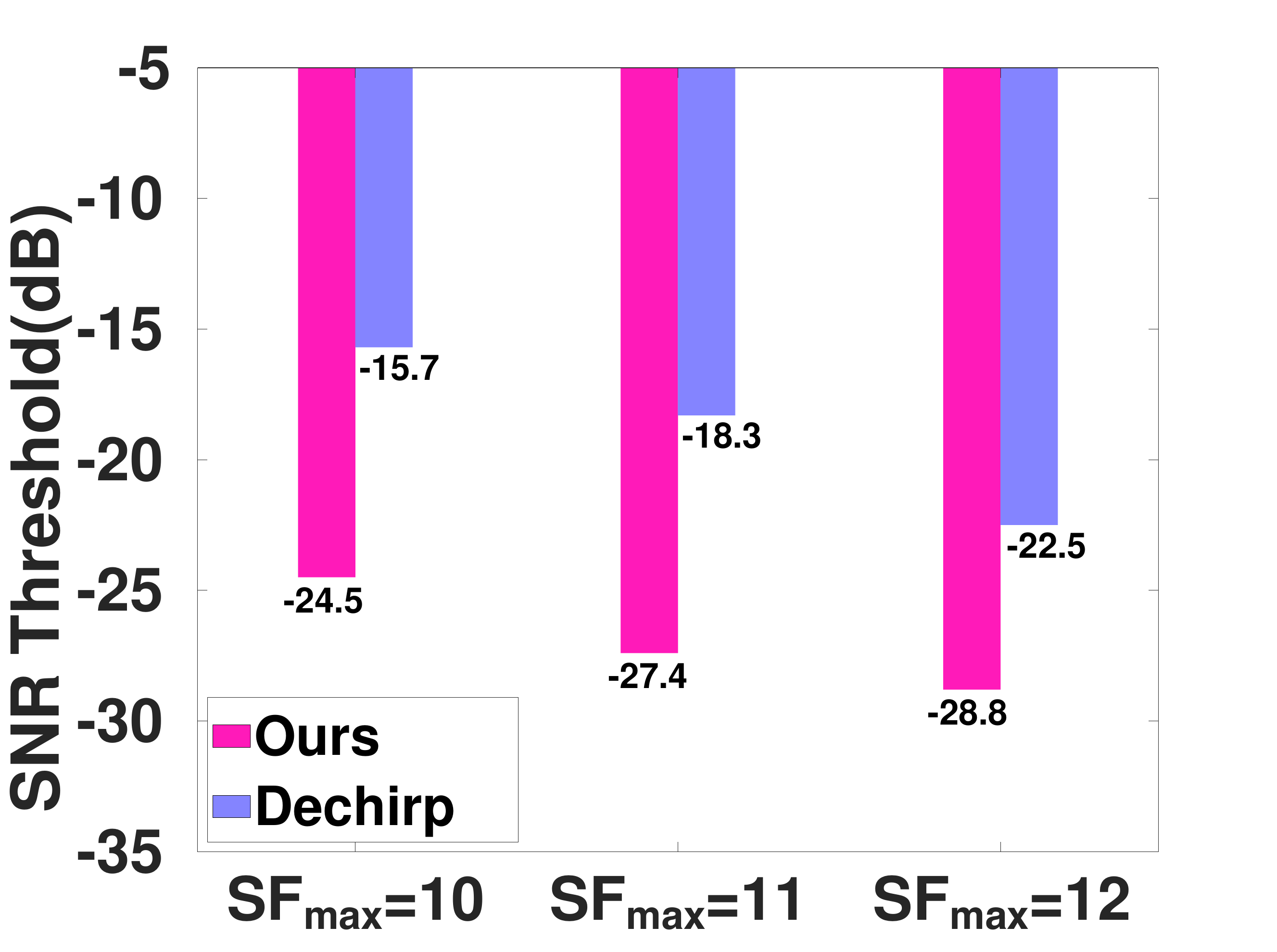}\label{fig:overall-performance-4}}
    \caption{In-domain performance of \ours.}
    \label{fig:overall-performance-indoor}
    \vspace{-5mm}
\end{figure}

\begin{figure*}[t]
    \centering
    \subfloat[Cross-env SER Distribution]{\includegraphics[width=0.25\textwidth]{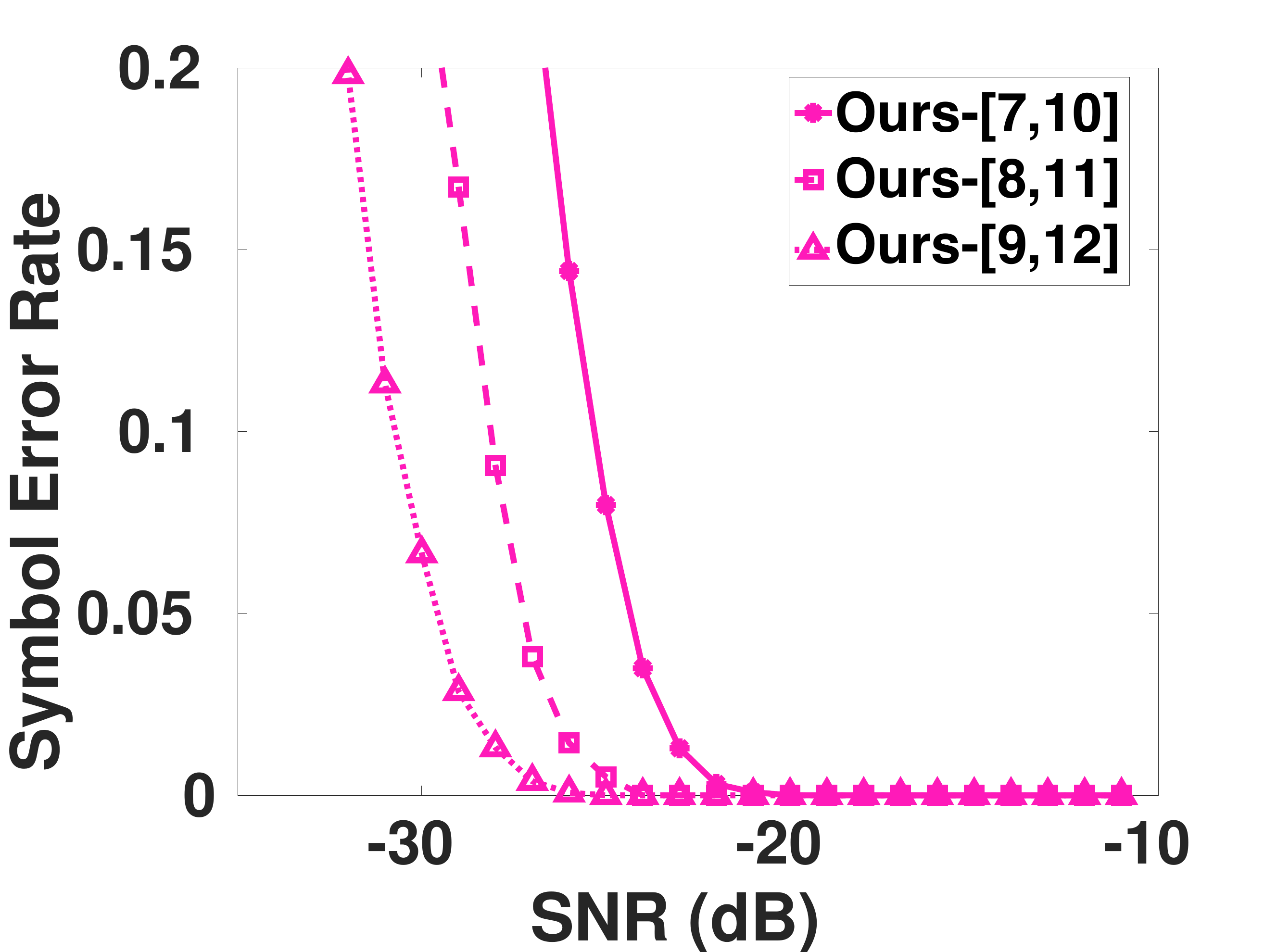}\label{fig:overall-performance-2}}
    \subfloat[Cross-node SER Distribution]{\includegraphics[width=0.25\textwidth]{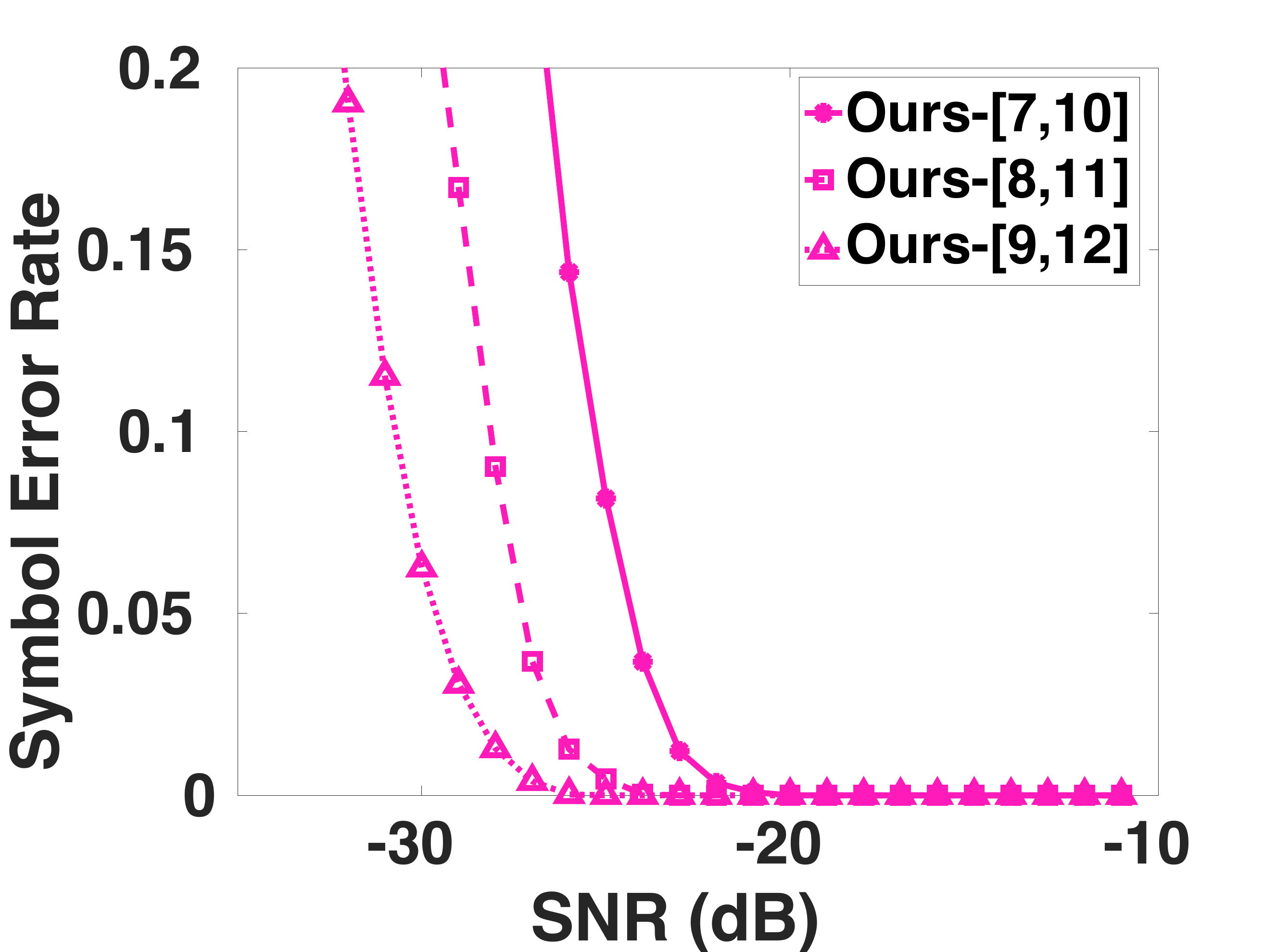}\label{fig:overall-performance-3}}
    \subfloat[Cross-env SNR Threshold]{\includegraphics[width=0.25\textwidth]{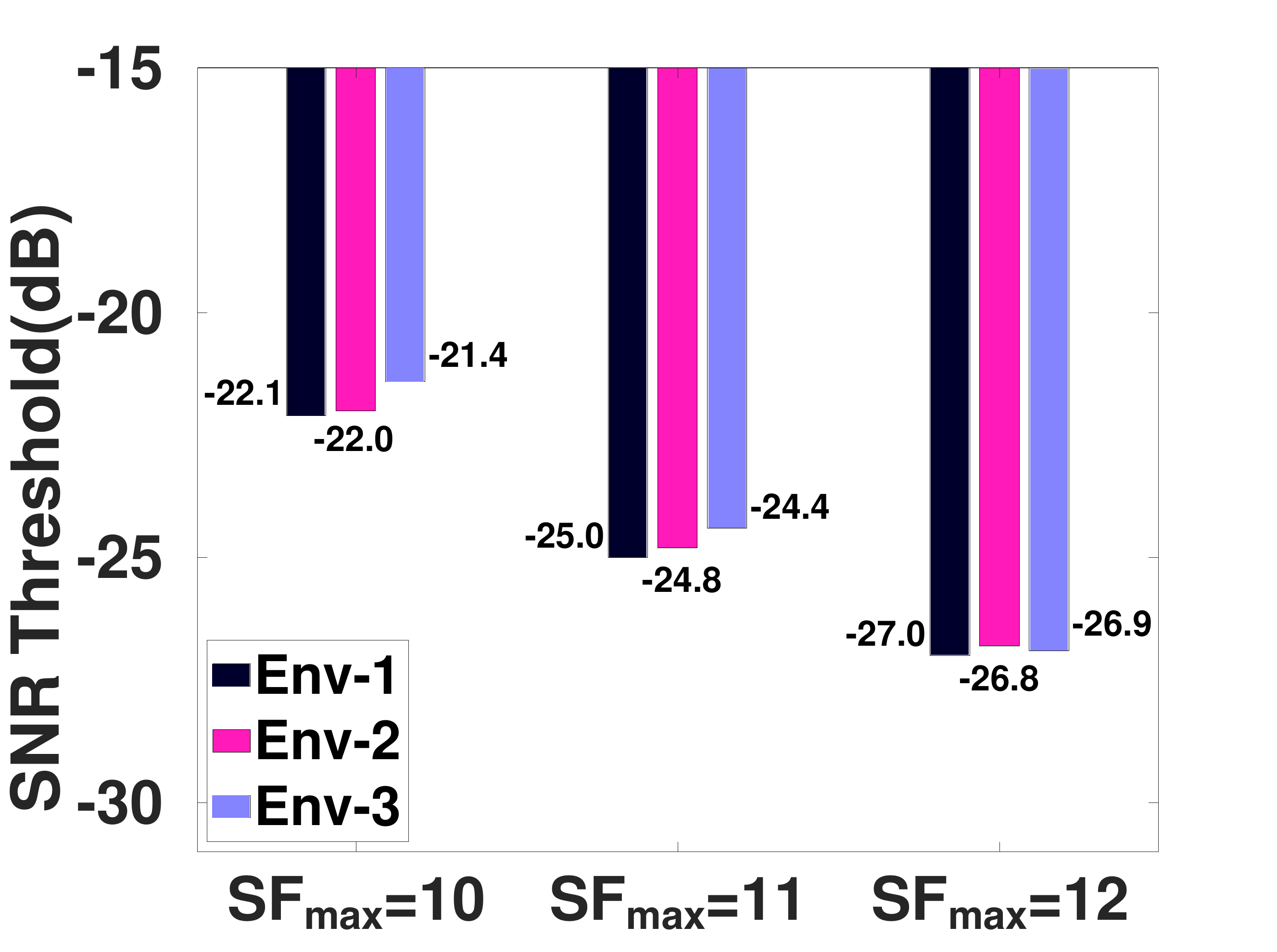}\label{fig:overall-performance-5}}
    \subfloat[Cross-node SNR Threshold]{\includegraphics[width=0.25\textwidth]{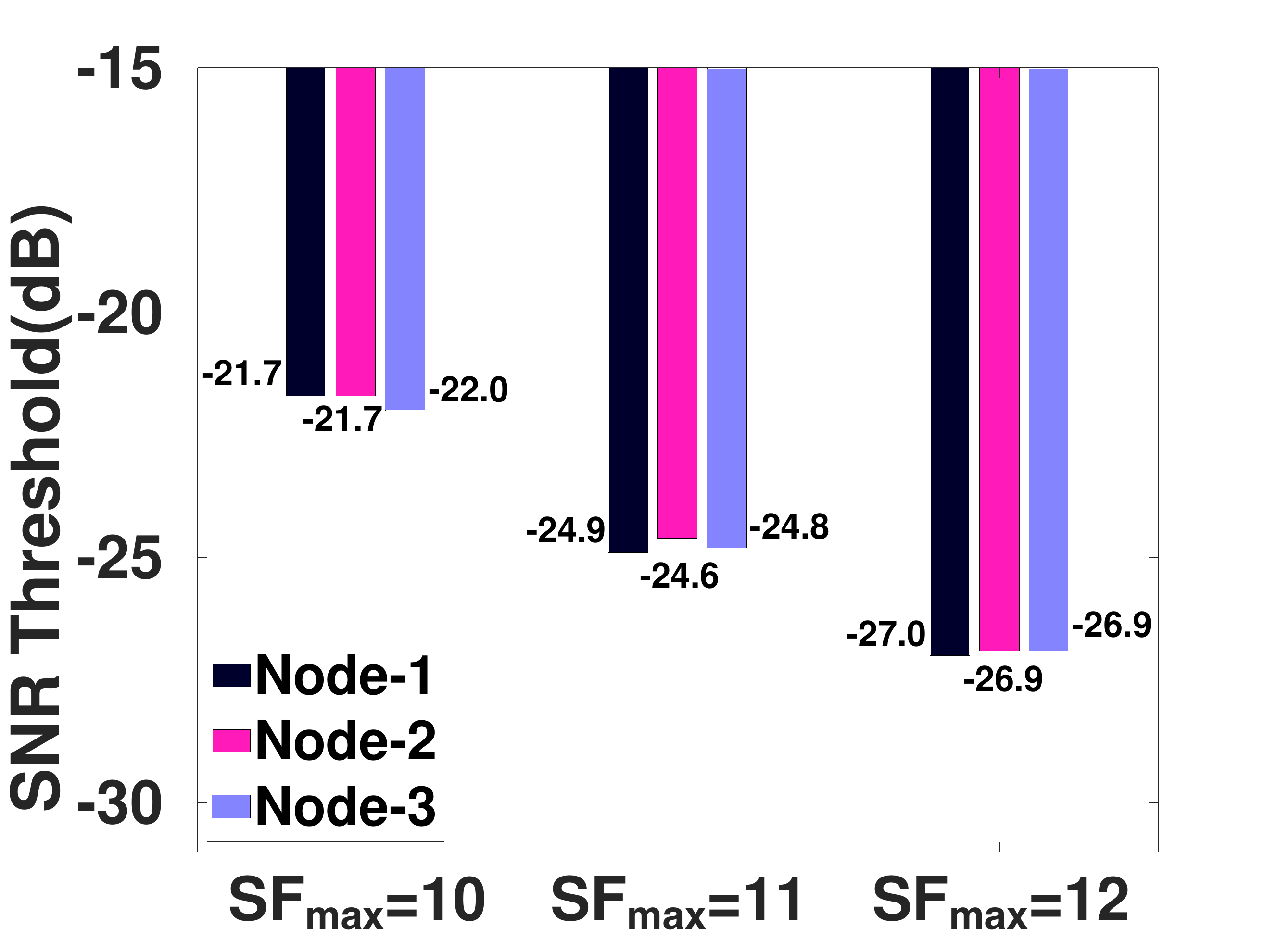}\label{fig:overall-performance-6}}
    \vspace{-1mm}
    \caption{In-domain and cross-domain performance of \ours on our indoor testbed.}
    \label{fig:overall-performance}
    \vspace{-3mm}
\end{figure*}

\subsection{Performance on Indoor Testbed}
\label{subsec-evaluation-1}

\noindent
\textbf{Setup:} We evaluate \ours with the in-domain testing data collected from our indoor testbed. To verify the SNR tolerance improvement given the same symbol period, we compare \ours, with energy and Dechirp under all three \ours configurations. We use the in-domain dataset.

\noindent
\textbf{Results:} Figure~\ref{fig:overall-performance-1} shows the SER distribution. 
We can observe that given the same symbol period (e.g. SF-10, SF-11, SF-12), \ours (pink line) has a consistently reduced SER in comparison with Dechirp (purple line). 
More importantly, for the encoding setting of SH-[7,10] and SH-[8,11], \ours achieves a comparable decoding performance with Dechirp using longer SF-11 and SF-12 chirps at low SNR conditions (e.g., $\leq$ 25~dB).
We further illustrate the SNR thresholds of different systems in Figure~\ref{fig:overall-performance-4}. 
In comparison with Dechirp, \ours improves the SNR threshold by 8.8~dB, 9.1~dB, and 6.3~dB for SF$_{\text{max}}$-10, SF$_{\text{max}}$-11, and SF$_{\text{max}}$-12, respectively. 
Note that the SNR threshold of \ours with low SF$_{\text{max}}$ encoding setting can still outperform LoRa-PHY with high SF chirp symbols. For example, compared to Dechirp with SF-12, the SNR threshold of \ours with SH-[7,10] encoding setting decreases from -22.5~dB to -24.5~dB (DNN decoder), respectively. This indicates \ours successfully extend LoRa-PHY to SF-12 or beyond.

\subsection{System Generalization Analysis}
\label{subsec-evaluation-3}

\noindent
\textbf{Setup:} To evaluate the effectiveness of our ``one-fits-all'' training method design, we directly deploy our pre-trained neural-enhanced decoder in new scenarios. The diversities come from three nodes (denoted as Node-1, Node-2, and Node-3) which are deployed in three indoor environments (denoted as Env-1, Env-2, and Env-3). We apply \ours to the data collected in cross-environment and cross-node manners. In a cross-environment manner, we group the signals from different nodes in the same environment to form a testing dataset. Similarly, in a cross-node manner, the signals from the same node in different environments belong to an identical testing dataset.

\noindent
\textbf{Results:} Figure~\ref{fig:overall-performance-2} to~\ref{fig:overall-performance-3} illustrate \ours SER distribution with all data in cross-environment and cross-node scenarios. In comparison to the corresponding SER distribution in the in-domain scenario (Figure~\ref{fig:overall-performance-1}), the SER increases slightly faster with the SNR decreasing under all three encoding configurations.
We further compute the SNR threshold for each cross-environment and cross-node scenario in Figure~\ref{fig:overall-performance-5} to~\ref{fig:overall-performance-6}. 
In comparison with the in-domain SNR threshold, \ours gets an SNR reduction of 2.4~dB on average for all cross-environment and cross-node scenarios without data re-collection and model re-training.
The reason is the noise diversity in a different environment, resulting in the pattern of a few symbols, which is hardly captured by our data augmentation.
We can resolve it with a fine-tuning process, which manually adjusts the initial frequency of the outlier symbols, then makes it agnostic to dynamic offsets in various periods, nodes, or environments.

\begin{figure}[!t]
    \centering
    \includegraphics[width=0.4\textwidth]{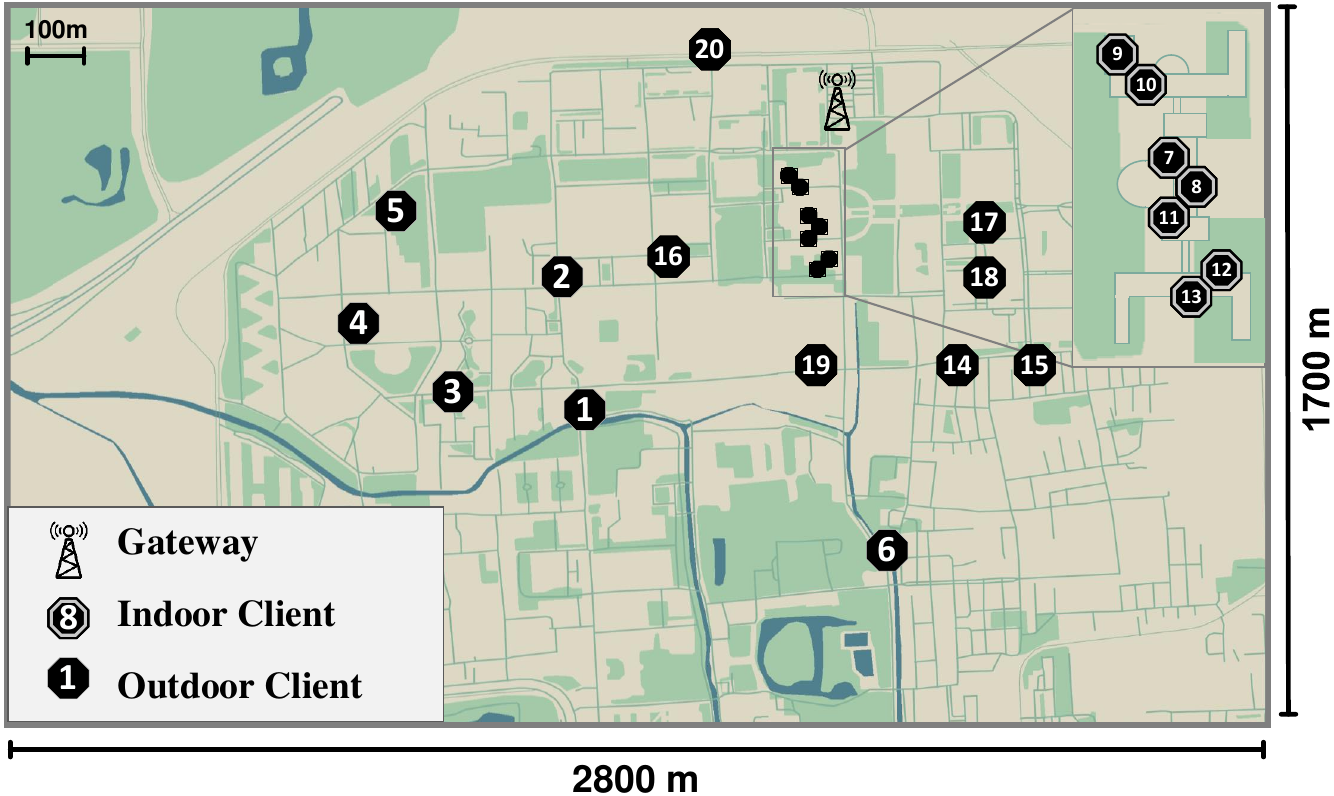}
    \vspace{-2mm}
    \caption{The campus-scale testbed with 20 LoRa nodes at 20 indoor and outdoor locations.}
    \label{fig:outdoor_deployment}
    \vspace{-4mm}
\end{figure}

\subsection{Performance on Campus-scale Testbed}
\label{subsec-evaluation-4}

\noindent
\textbf{Setup:} \red{SH is adopted by those LoRa nodes experiencing weak links to improve network coverage. 
The SH encoder extends its range and reduces the SNR threshold by lowering the data rate and neural-enhanced decoder. We conduct campus-scale experiments to verify the gain of SH encoder-decoder co-design with extremely low SNR conditions.} Figure~\ref{fig:outdoor_deployment} illustrates the deployment of our outdoor testbed at a campus ($2800m\times1700m$). We randomly deploy 20 COTS LoRa nodes at 20 different NLOS positions covering indoor and outdoor scenarios. For indoor nodes, concrete walls are the main obstacles. Buildings and trees are the obstacles in the outdoor environment. For each LoRa node at a position, we collect tens of packets, and each contains 40 payload symbols. We repeat the data collection with three encoding methods of the standard LoRa, \ours using SH-[9,12] configuration, and Ostinato using \textbf{Ostinato-4} configuration, respectively.
We first evaluate the preamble detection accuracy of our design compared to the standard LoRa. In addition, for those detected packets, we compare the decoding SER between \ours and Ostinato.
For our SH-[9,12], we apply the DNN-based decoder pre-trained with our synthesis dataset collected in $\S$~\ref{subsec-evaluation-2} on the weak signal packets.

\begin{figure}[!t]
    \centering
    \subfloat[Cross-position Performance]{
    \label{subfig-detect-accuracy}
    \includegraphics[width=0.25\textwidth]{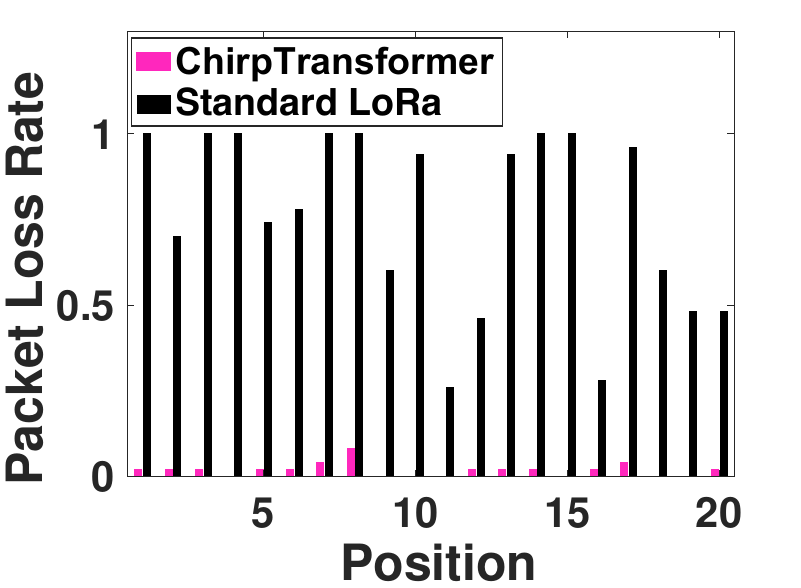}
    }
    \subfloat[CDF Performance]{
    \label{subfig-detect-cdf}
    \includegraphics[width=0.24\textwidth]{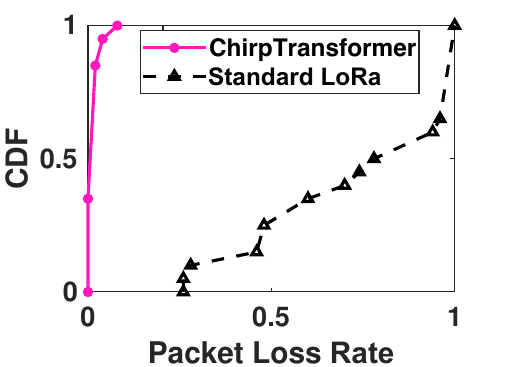}
    }
    \vspace{-2mm}
    \caption{The packet detection accuracy on campus-scale testbed with extremely-low SNR.}
    \label{fig:detection}
   \vspace{-3mm}
\end{figure}

\noindent
\textbf{Results for packet detection:} To understand whether a \ours packet can be successfully detected at extremely low SNR conditions, we evaluate \ours's performance on packet detection for all 20 positions. 
Figure~\ref{subfig-detect-accuracy} shows the comparison of the packet loss rate between \ours and the standard LoRa. Figure~\ref{subfig-detect-cdf} shows the CDFs of the packet loss rate for \ours and the standard LoRa. 
We can see that the standard LoRa suffers a high packet loss rate at all positions, where more than 90\% of packets are undetected at 10 out of 20 positions. In comparison, \ours achieves a lower than 6\% packet loss rate at all positions.
This is because the standard LoRa detects LoRa preambles by searching for continuous identical frequency-domain energy peaks by applying the dechirp to the continuously received signals. Thus, it requires the energy peak of each chirp symbol in the preamble to be detectable. However, the real SNR at most of the positions is much lower than the SF-12 SNR threshold (e.g., -22.4~dB), leading to undetectable energy peaks. \red{After mitigating the CFO and SFO in a preamble, \ours concentrates the energy of all preamble chirp symbols to detect a LoRa packet. The results verify that \ours can reliably detect any LoRa packets at the extremely low SNR.}

\begin{figure}[!t]
    \centering
    \subfloat[Cross-position Performance]{
    \label{subfig-campus-ser}
    \includegraphics[width=0.25\textwidth]{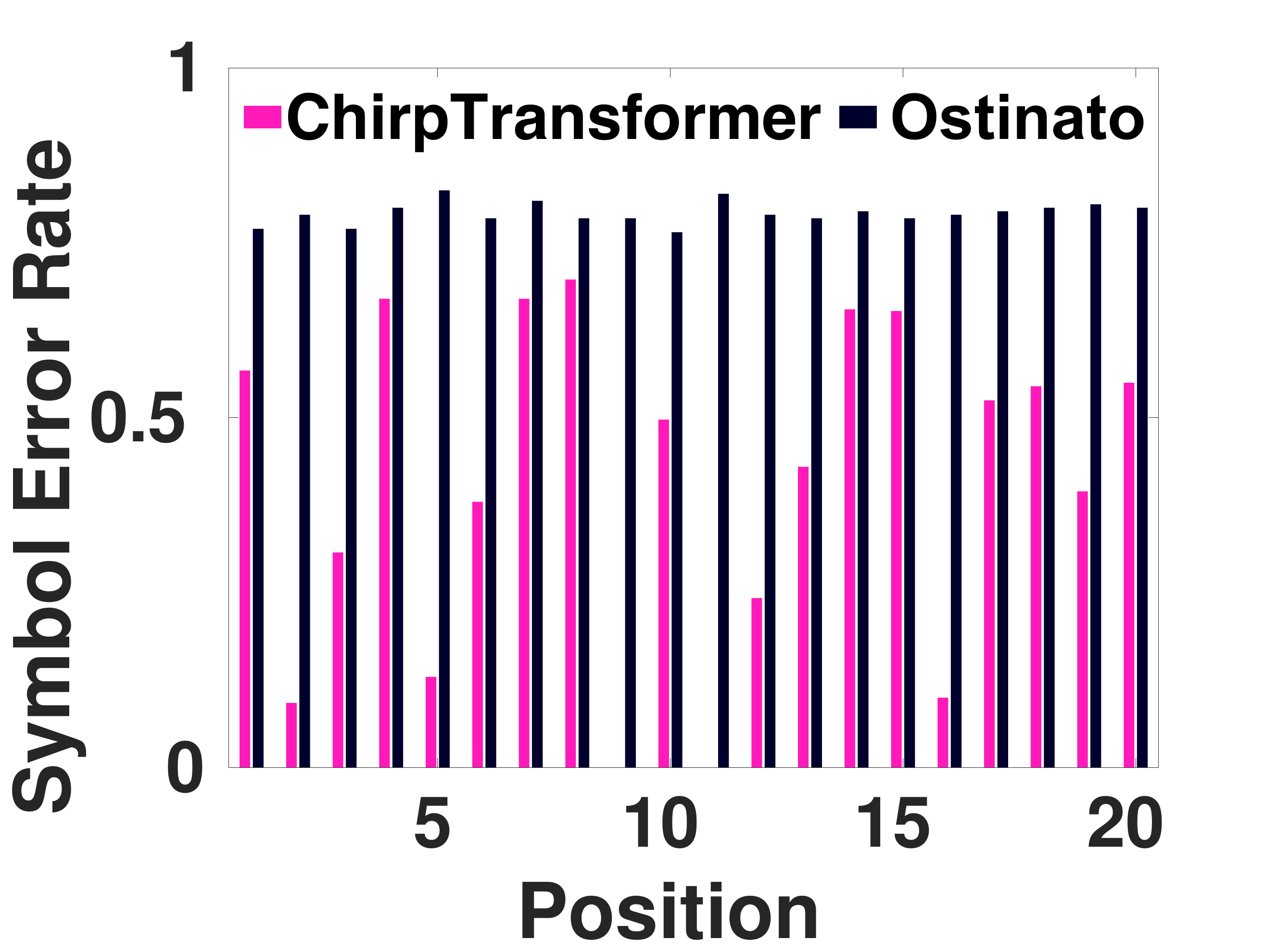}
    }
    \subfloat[CDF Performance]{
    \label{subfig-campus-ser-cdf}
    \includegraphics[width=0.24\textwidth]{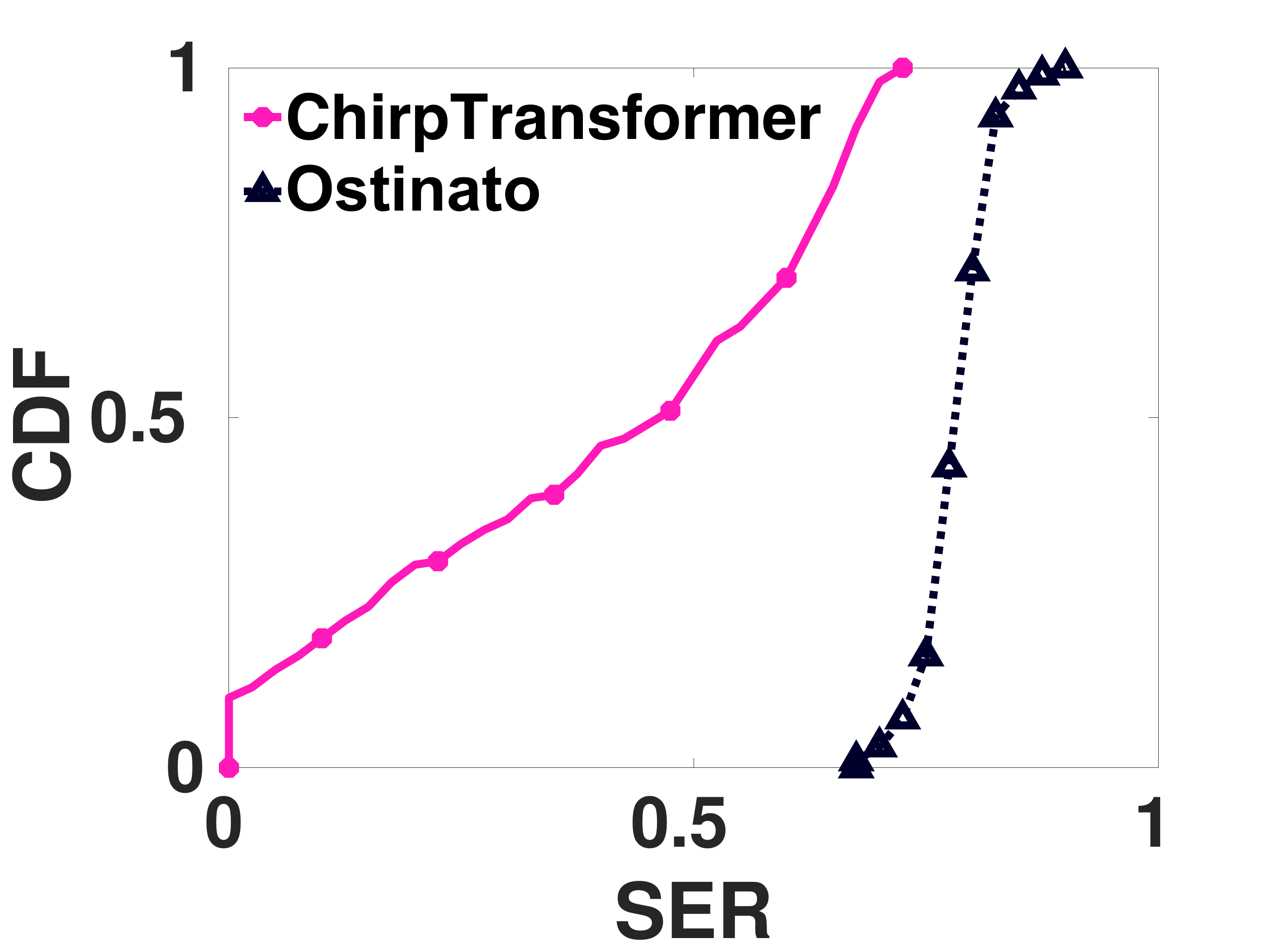}
    }
    \vspace{-3mm}
    \caption{The comparison of communication reliability on campus-scale testbed.}
    \label{fig:campus}
    \vspace{-3mm}
\end{figure}

\noindent
\textbf{Results for decoding SER:} We utilize our preamble detection design to detect Ostinato packets. Then, we compare \ours with Ostinato by computing the decoding SER for the detected symbols at all 20 positions.
As shown in Figure~\ref{subfig-campus-ser-cdf}, we can see the median SER of Ostinato is about 79.25\%. In comparison, \ours can achieve a much lower median SER of 40.54\%, indicating the boundary of the communication range will be enlarged greatly. 
The reason is that Ostinato suffers from severe noises in the wild environment, leading to the signals not being coherently combined. At the same time, our neural-enhanced decoder can tolerate it by exploring multi-dimensional features.
The specific SER across 20 positions is shown in Figure~\ref{subfig-campus-ser}. The SER of \ours is significantly lower than Ostinato, which exceeds $75\%$ at all 20 positions. The decoding results of Ostinato appear random, leading to consistently high and unstable symbol error rates across 20 locations.
In addition, we observe that \ours has a higher SER in some positions, such as positions 4, 7, 8, 14, and 15, than others.
The reason is that some coexisting wireless interference also brings new noise patterns that the pre-trained DNN model does not see, degrading the SER.
We can involve an online fine-tuning process to deal with those new noise patterns~\cite{li2021nelora}

\section{Related Work}
\label{sec-related-work}

\noindent
\textbf{LoRa Reliable Preamble Detection:} Some works~\cite{li2021nelora, hou2022malora} coherently combine the duplicate chirps in the preamble of a LoRa packet to detect it at extremely-low SNR conditions. A similar idea can be adopted by an encoder that uses several repeated chirps to form a symbol for achieving reliable communication at extremely-low SNR. However, such an encoder design is neural-inefficient. Thus a long-period symbol is hard to be decoded with a neural-enhanced decoder providing extra SNR gains. In contrast, \ours's encoder is neural-efficient since it lowers the encoded data bits per symbol to tolerate extremely-low SNR instead of increasing the symbol period. Therefore, the encoder-decoder co-design of \ours maximizes the SNR threshold and extends the applicability of LoRa communication under complex NLoS scenarios.

\blue{
\noindent
\textbf{LoRa Reliable Decoding:} 
Demeter~\cite{ren2024demeter} proposes a hardware solution to align the polarization to improve signal reception.
XCopy~\cite{xia2023xcopy} coherently combines
retransmitted packets to improve SNR.
LoRaTrimmer~\cite{du2024loratrimmer} trims FFT range  to remove the noise and merge the power.
Recent work~\cite{xiaolong2025icdcs, yu2024resolve, xu2023slora} propose how to handle with cross-channel interference for LoRa decoding. 
Previous research~\cite{yu2024fdlora,yu2024revolutionizing, hu2020sclora,yu2023enabling} focus concurrent transmission problems
By utilizing either multiple gateways and LoRa nodes, recent studies~\cite{dongare_charm_2018,balanuta_cloud-optimized_2020,eletreby_empowering_2017, hou2022malora} bring extra SNR gains.
Charm~\cite{dongare_charm_2018} coordinates multiple gateways to decode weak signals by detecting the combined energy peak in the spectrum. 
OPR~\cite{balanuta_cloud-optimized_2020} provides an increasing gain in bit error recovery as a function of the number of feasible gateways. 
Choir~\cite{eletreby_empowering_2017} exploits the correlation across co-located LoRa nodes, enabling a larger communication range than an individual one.
Chime~\cite{gadre_frequency_nodate} uses multiple gateways to eliminate the multi-path interference by frequency selection to capture extra SNR gains for LoRa transmissions. 
MALoRa~\cite{hou2022malora} utilizes external clock oscillator to mitigate hardware offset for multi-antennas.
In contrast, we fully exploit the SNR gains using an encoder and neural-enhanced decoder co-design with low infrastructure cost. And \ours works for beyond SF-12 ultra-weak signal decoding and provide more SNR gain than previous methods. In addition, with signals recorded by different gateways, \ours's SNR gains can be further enhanced by the diversity.
}

\noindent
\textbf{AI-augmented LoRa Systems:} 
Deep learning has been broadly adopted to augment various wireless applications~\cite{AIWirelessSensingSurvey_CSUR20} such as pose estimation~\cite{zhao2018rf,adib2015capturing,WiPose}, human monitoring detection~\cite{ha2021wistress,zhao2016emotion}, and gesture recognition~\cite{Widar3.0,WiHF}. DeepLoRa~\cite{deepLora} uses bidirectional LSTM to estimate the path loss of a LoRa link accurately. 
DeepSense~\cite{chan_deepsense_2019} uses DNN to detect LoRa signal from coexisting interference to enable efficient carrier sense.
NELoRa~\cite{li2021nelora,du2023nelora} develops a DNN demodulator capturing multi-dimension features to obtain 1.84-2.35~dB SNR gains but suffers from exhausted computation and training overhead when the SF is getting larger. SRLoRa~\cite{du2023srlora} further optimizes the DNN demodulator for multiple gateways. In comparison with NELoRa, based on our neural-efficient SF-configuration-based encoding, we develop a customized DNN decoder to improve the computation efficiency by more than 3$\times$.

\blue{
\section{Discussion and Future Work}
\noindent
\textbf{Compatible Gateway:}
While our technique is compatible with COTS LoRa nodes, our decoder algorithm currently lacks support for existing COTS gateways. This limitation can be addressed by offloading computation to the cloud or implementing our design on next-generation gateways with enhanced processing capabilities.

\noindent
\textbf{Higher SNR Tolerance:}
Future work could improve both encoder and decoder designs. For the encoder, employing a more diverse symbol hopping pattern with additional SF combinations and varying initial frequency offsets could be beneficial. For the decoder, more efficient NN denoising model is also a promising direction.

\noindent
\textbf{Mobile Scenarios:}
In mobile scenarios, the limited training samples still make few-shot learning challenging. Expanding data sample diversity and employing large-scale generative AI-based data augmentation can better adapt to various mobile conditions and enhance the decoder's robustness. For the mobile sensing scenario~\cite{hydra24liu,Adonis25Liu,gan2023poster,Proteus25Liu}, recent multimodal and neural‑enhanced mmWave sensing applications demonstrate how limited training datasets can be robustly expanded and performance improved under dynamic conditions. For high-speed scenarios like satellite networking~\cite{ren2024sateriot, gadre2024adapting}, both standard LoRa and \ours face Doppler challenges that require further research in the future.

}

\section{Conclusion}
\label{sec-discussion-conclusion}

We have witnessed the dramatic LoRa communication distance decrease due to the severe signal attenuation in NLoS scenarios. In this paper, we propose \ours, a LoRa encoder and decoder co-design to enable the reliable connection between a LoRa node and gateways at extremely-low SNR (e.g., -28.8~dB). 
At the encoder side, instead of only using the frequency-domain energy feature to encode chirps, \ours incorporates an SF-configuration-based coding method to exploit the different time-domain spectrogram patterns among the chirps with different SFs. The encoder can be implemented on COTS LoRa nodes and mimics the standard encoding with larger SFs beyond SF-12.
At the decoder side, we adopt the idea of a neural-enhanced decoder to obtain extra SNR gains by exploring multi-dimensional features in the spectrograms generated by our encoder design.
We further customize the DNN model by adjusting the input size, replacing structure layers, and compressing the model to improve its efficiency, reliability, and generalizability.
We have conducted extensive experiments in both indoor and outdoor testbeds to evaluate the performance of \ours.
\ours enables successful data decoding even when the SNR approaches -28.8~dB, which significantly enhances communication reliability.
The computation efficiency of our DNN decoder is $3.14\times$ higher than NELoRa~\cite{li2021nelora}.
%


\bibliographystyle{IEEEtran}
\bibliography{reference}

\end{document}